\documentclass[sn-standardnature]{sn-jnl}

\jyear{2023}%

\usepackage[utf8]{inputenc}
\usepackage{graphicx}
\usepackage{graphics}
\usepackage{tabularx}
\usepackage{bm}
\usepackage{cleveref}
\usepackage{caption}
\usepackage{xr}

\usepackage{adjustbox}

\usepackage{color, soul}
\usepackage{gensymb}

\graphicspath{{./}}

\usepackage[superscript,biblabel]{cite}

\usepackage[superscript,biblabel]{cite}

\newcommand{\beginsupplement}{%
        \setcounter{table}{0}
        \renewcommand{\thetable}{S\arabic{table}}%
        \setcounter{figure}{0}
        \renewcommand{\thefigure}{S\arabic{figure}}%
        \setcounter{page}{1}
        \renewcommand{\thepage}{S\arabic{page}}%
     }

\begin{document}

\title[]{Creation of crystal structure reproducing X-ray diffraction pattern without using database}

\author*[1]{\fnm{Joohwi} \sur{Lee}}\email{j-lee@mosk.tytlabs.co.jp}
\author[1]{\fnm{Junpei} \sur{Oba}}
\author[1]{\fnm{Nobuko} \sur{Ohba}}
\author[1]{\fnm{Seiji} \sur{Kajita}}
\affil[1]{\orgdiv{Toyota Central R\&D Labs., Inc.},  \orgaddress{\street{41-1 Yokomichi}, \city{Nagakute, Aichi}, \postcode{480-1192}, \country{Japan}}}

\abstract
{When a sample's X-ray diffraction pattern (XRD) is measured, the corresponding crystal structure is usually determined by searching for similar XRD patterns in the database. However, if a similar XRD pattern is not found, it is tremendously laborious to identify the crystal structure even for experts.
This case commonly happens when researchers develop novel and complex materials.
In this study, we propose a crystal structure creation scheme that reproduces a given XRD pattern.
We employed a combinatorial inverse design method using an evolutionary algorithm and crystal morphing (Evolv\&Morph) supported by Bayesian optimization, which maximizes the similarity of the XRD patterns between target one and those of the created crystal structures.

For sixteen different crystal structure systems with twelve simulated and four powder target XRD patterns, Evolv\&Morph successfully created crystal structures with the same XRD pattern as the target (cosine similarity $>$ 99\% for the simulated ones and $>$ 96\% the experimentally-measured ones). 

Furthermore, the present method has merits in that it is an automated crystal structure creation scheme, not dependent on a database.
We believe that Evolv\&Morph can be applied not only to determine crystal structures but also to design materials for specific properties.
}

\keywords{crystal structure creation, evolutionary algorithm, crystal morphing, Evolv\&Morph, XRD similarity, inverse design}

\maketitle
\newpage
\section*{INTRODUCTION}\label{sec1}

When synthesizing a material in a particular composition, one wants to confirm whether the intended crystal structure is successfully synthesized as a crystalline phase. 
An XRD (X-ray diffraction) analysis is used to determine atomistic and molecular structures \cite{XRD1967}. 
It has a wide range of applications, including determination of crystalline phase, orientation, lattice parameters, and grain size. 
Furthermore, because XRD is prevalent to market and relatively easy to handle, the method can be said to be the first analysis in investigation of the crystal structure and phase.

For determining the crystal structure based on the XRD analysis, the material database (DB) is widely used together. 
There are big material DB including more than hundreds of thousands of XRD patterns and corresponding crystal structures, such as Powder Diffraction File (PDF\textsuperscript{\texttrademark})\cite{ICDD-PDF} produced/managed by International Centre for Diffraction Data (ICDD\textsuperscript{\textregistered}) and Inorganic Crystal Structure Database (ICSD) \cite{ICSD2002}.
XRD patterns can be directly simulated from crystal structures saved in the DB.
Conventionally, a crystal structure of a measured sample is identified by matching its XRD pattern with those of candidate materials searched in DB.

Recently, various methods have been suggested for more correct and efficient identification of material systems or crystal structures based on the XRD analysis and DB. 
Machine learning models \cite{Park2017,Oviedo2019,Suzuki2020} were proposed to predict crystal systems and space groups by inputting XRD patterns. 
Griesemer \textit{et al}. suggested a prototype search method based on exploiting DB and first-principles calculations to identify the structures of approximately five hundred compounds from experimental XRD patterns missed from DB \cite{Griesemer2021}. 
Dong \textit{et al}. constructed a deep learning model to predict XRD pattern only by the input of chemical composition directly \cite{deepXRD}. 
However, when the measured XRD pattern indicates an unknown crystal structure, it is often impossible to find similar XRD patterns in DB. 
Because such advanced methods strongly depend on the accumulated DB, the generated crystal structure is not guaranteed to match the measured XRD pattern successfully. 

To reduce the gap between the measured and candidate XRD patterns, a Rietveld refinement \cite{Rietveld}, which can directly tune a candidate crystal structure to approach a similar XRD pattern, is usually employed.
Because this method needs to optimize complex combinations of various parameters, high expertise is needed to successfully reduce the difference between the XRD patterns. 
Recently, Ozaki \textit{et al}. proposed Black-Box Optimization Rietveld (BBO-Rietveld) method \cite{BBO-rietveld} that automatically optimizes various combinations of parameters for Rietveld refinement. 
BBO-Rietveld is easier to be handled even by nonexperts, providing a higher success probability of Rietveld Refinement. 
However, Rietveld refinements strongly depend on an initial structure for the optimization.
If an XRD pattern of the initial structure loaded from DB differs significantly from the measured XRD pattern, the Rietveld refinement often does not succeed. 
Therefore, it is desired to develop an inverse design\cite{inversedesign,inversedesign2} method to directly create crystal structure reproducing the target XRD pattern without relying on any DB.

In this study, we propose a scheme that consists of evolutionary algorithm \cite{USPEX1,Falls2020} and cyrstal morphing \cite{morphing} (Evolv\&Morph) for the direct creation of crystal structures similar to a target XRD pattern. 
Evolv\&Morph does not use prior knowledge in crystal structures such as structural DB. 
We show that Evolv\&Morph successfully created structures reproducing the target XRD patterns for sixteen different material systems.

\section*{RESULTS}\label{sec2}
\subsection*{Overview of the present scheme}\label{subsec1}
\begin{figure*}[htp]
 \begin{center}
  \includegraphics[width=0.9\linewidth]{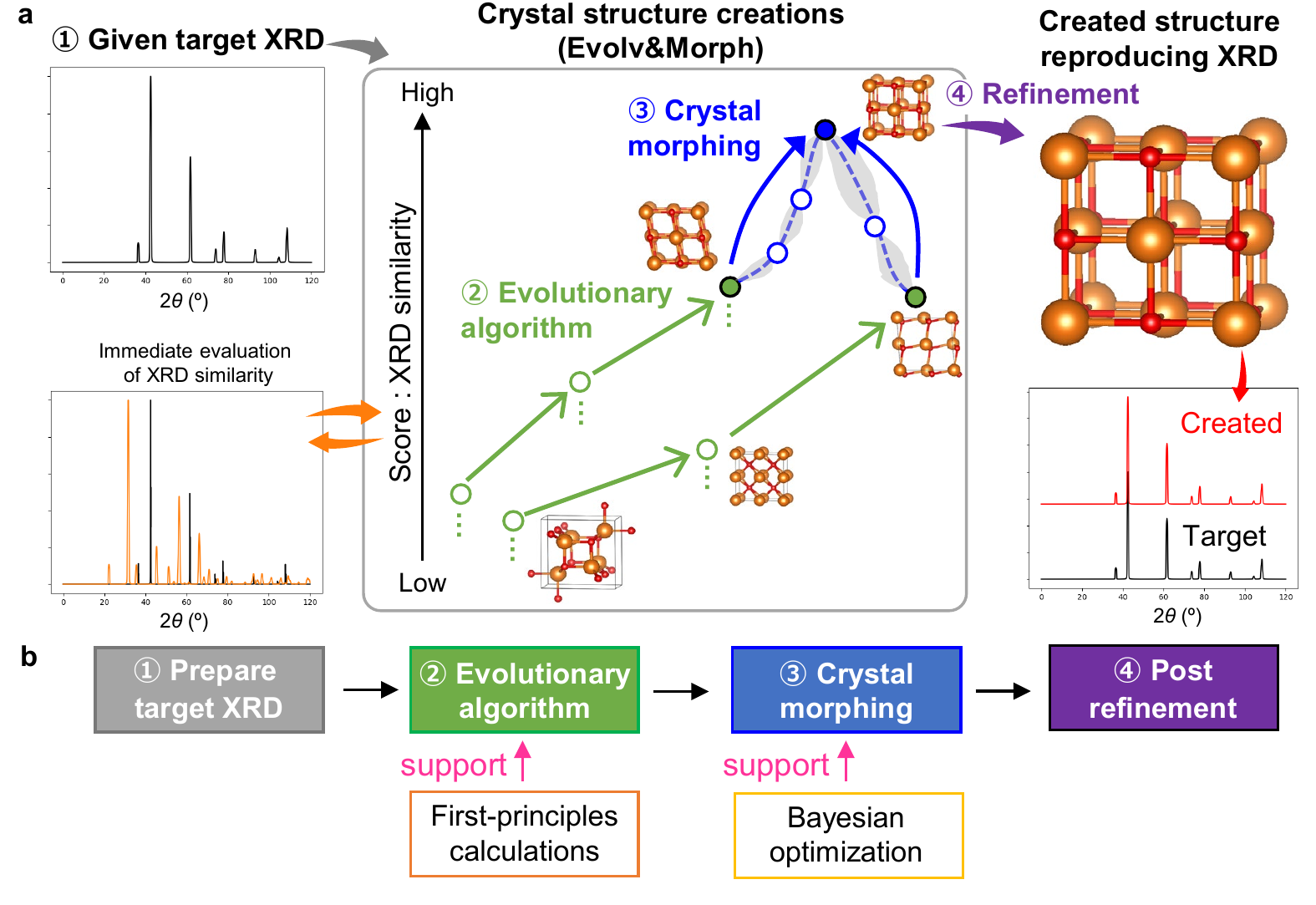}
 \caption{
  \textbf{Overview to create a crystal structure reproduces the target XRD pattern.} 
  \textbf{a} An overview.
  Circled numbers indicate the sequence of the procedure.
  First, a target XRD pattern for reproducing is given. 
  Evolv\&Morph, consisting of the evolutionary algorithm and crystal morphing, tries to create various crystal structures with optimizing the target score: similarity of their XRD patterns with respect to the target. 
  In the center box, the open circles indicate some of the created structures, and the closed circles indicate the structure with the highest similarity score for each method. 
  The optimization target score is evaluated immediately during the creation of the structures.
  Finally, after refinement, the structure with almost the same XRD pattern as the target is suggested.
  \textbf{b} A simple workflow chart. Circled numbers also indicate the sequence of the procedure. 
  Note that the first-principles calculations can be added to optimize the structures created by the evolutionary algorithm to make them more stable ones.
  Bayesian optimization provides the crystal morphing with an optimization function by guiding what structure should be created by the morphing.   
  For the crystal morphing procedure in \textbf{a}, the dashed curve and gray shaded area indicate the predictive mean and confidence interval of Bayesian optimization, respectively.
 }
\label{fig:overview}
\end{center}
\end{figure*}

Figure \ref{fig:overview} shows an overview of Evolv\&Morph. 
The goal is to create a crystal structure with the same XRD pattern as a given target. 
To achieve the goal, the main part of Evolv\&Morph requires two important factors. 
One is to create an enormous number of structures automatically, 
and another is to select and modify such structures to maximize (optimize) their similarity score of the XRD pattern with respect to the target.
To this end, the similarity score is required to be evaluated immediately.
As structure creation methods, we employed evolutionary algorithm \cite{USPEX1,Falls2020} and crystal morphing \cite{morphing}. 

The evolutionary algorithm is a heuristic optimization method for creating various crystal structures. 
It has been widely used to suggest novel structures that have optimized target property (fitness) of thermodynamic energy \cite{USPEX1,Falls2020}, or other more practical properties such as defect formation energy \cite{Lee2018} and hardness \cite{Oganov2010}. 
We chose a similarity score of XRD patterns to find a structure possessing a target XRD pattern as the optimized target property. 
It tries to create structures based on various genetic operators to maximize the similarity score. 
In addition, the first-principle calculations optimize each structure created by evolutionary algorithm to become a more stable one.

Crystal morphing generates intermediate crystal structures between given two structures. \cite{morphing} 
As an application example, several virtual geometric structures with four carbon atoms had been morphed in such a way as to have the target XRD patterns. 
For practical material systems, it was not yet confirmed whether such an application of crystal morphing is successful or not in offering crystal structure reproducing XRD patterns.
Its optimization of similarity score can be supported by external optimization functions such as Bayesian optimization \cite{bayesian}. 

In this study, when a crystal structure reproducing a target XRD pattern is searched, evolutionary algorithm was performed five times to create structures with each different highest similarity score of XRD pattern with respect to the target. 
Then, crystal morphing was followed using the input structures obtained from the different evolutionary algorithms.

Among the various created structures, the structures with significantly high similarity scores can be refined by post–process such as Rietveld refinement and symmetrization. 
Such refinement methods further slightly tune the structures to increase the similarity score. 
Finally, the structures with the highest similarity scores can be recommended as candidates for reproducing the target XRD pattern successfully.
More details of crystal structure creation methods (evolutionary algorithm and crystal morphing), their supporting methods (first-principles calculations and Bayesian optimizations, respectively), and post refinements are written in \nameref{secmethod} section.

\subsection*{XRD simulations and similarity metrics}\label{subsec5}
When a crystal structure is given, the XRD patterns can be simulated immediately.
Accordingly, the simulated XRD patterns of numerous created structures were also immediately created.

For the target XRD patterns for reproducing, simulated XRD patterns from twelve crystal structures in ICSD were employed. Herein, the structures for reproducing the target XRD patterns are referred to as target structures.
Six binary and six ternary compound systems were selected with each cubic, hexagonal, trigonal, tetragonal, orthorhombic, and monoclinic structure. They are listed in Table \ref{tab:main}.
Some typical structures were selected (for example, MgO in the Rocksalt structure and Al$_2$O$_3$ in the Corundum structure),
while the others were randomly chosen from the material list. 
In ICSD, the lattice parameters and internal coordinates of constituent atoms have primarily been obtained through synthesis and measurement.
However, creating a crystal structure reproducing the experimentally measured XRD pattern is more challenging because it includes other effects not included in simulations, such as background noises and instrumental artifacts. 
Therefore, in addition, four powder XRD patterns in RRUFF mineral database\cite{RRUFF,RRUFF-site} were also employed as the target.

As a similarity metric for two XRD patterns, a cosine similarity (\textit{S}$_\textrm{cos}$) was employed. 
Simply, a higher score indicates that two vectors \textit{\textbf{x}} and \textit{\textbf{y}} of XRD patterns have more overlapped peaks. 
They have the same number of bins (the same $2\theta$ range).
The maximum $S_\textrm{cos}$ is 100\%, which means that two XRD patterns are totally identical.
$S_\textrm{cos}$ can be obtained from the following equation:
\begin{eqnarray}
S_{\cos} (\bm{x},\bm{y})= \frac{\bm{x} \cdot \bm{y}}{\|\bm{x}\|\|\bm{y}\|}
= \frac{\sum_i x_i y_i}{\sqrt{\sum_i x_i^2} \sqrt{\sum_i y_i^2}}.
\end{eqnarray}
where $\|$ $\|$ is a $L$2-norm, $i$ is an index of bin ($2\theta$), and $\cdot$ is a dot product. 
Two XRD patterns were smeared with a 0.5°-width Gaussian before calculating $S_\textrm{cos}$.

Note that $S_\textrm{cos}$ was set at the maximum score obtained after isotropic volume changes to complement a critical weak point of this metric: 
it is too sensitive to the peak shift \cite{similarity-compare,Cha2007}, which corresponds to the change of lattice volume (or parameter) of crystal structure. 
The exact method how to solve the problem is discussed in Supplementary Section \nameref{csdemerit} and Figs. \ref{fig:mgoxrd}–\ref{fig:cossimvol}.

\begin{table}[htp]
\begin{center}
\begin{minipage}{\textwidth}
\caption{
$S_\textrm{cos}$ obtained by crystal structure creation methods used in this study for twelve materials. The target XRD patterns were simulated using the crystal structures in ICSD. The values after $\pm$ indicate the standard deviation of $S_\textrm{cos}$ from five different trials of the evolutionary algorithm. The score in parenthesis [] indicates the highest one using the input structures obtained by the evolutionary algorithm without performing post crystal morphing.}\label{tab:main}

\begin{tabular*}{\textwidth}
{@{\extracolsep{\fill}}l@{\extracolsep{\fill}}cccccc@{\extracolsep{\fill}}}
\toprule%
\multicolumn{2}{@{}c@{}}{Target} & \multicolumn{4}{@{}c@{}}{Highest $S_\textrm{cos}$ of XRD pattern}\\
\multicolumn{2}{@{}c@{}}{~} & \multicolumn{4}{@{}c@{}}{ of created structure with respect to target (\%)}\\
\cmidrule{1-2}\cmidrule{3-6}
Materials \footnotemark[1] & Space & \multicolumn{2}{@{}c@{}}{By evolutionary} & By crystal & By Refine- \\
~ & group & \multicolumn{2}{@{}c@{}}{}{algorithm} & morphing \footnotemark[2] & ment \footnotemark[3] \\
\cmidrule{3-4}
~ & ~ & Mean & Max. & ~ & ~ & \\
\midrule%
MgO (4:4) & $Fm$–3$m$ & 99.5 $\pm$0.5 & 99.9 & ~ & 99.9 \\
NbCu$_3$Se$_4$ (1:3:4) & $P$–43$m$ & 99.2 $\pm$0.4 & 99.6 & ~ & 100 \\ 
GaN (2:2) & $P$6$_3mc$ & 98.6 $\pm$0.8 & 99.4 & ~ & 100 \\ 
Zr$_3$Cu$_4$Si$_2$ (3:4:2) & $P$–62$m$ & 90.4 $\pm$6.1 & 96.4 & 98.7 & 99.6 \\ 
~ & ~ & ~ & ~ & ~ & [99.6] \\ 
Al$_2$O$_3$ (4:6) & $R$–3$c$ & 98.5 $\pm$0.8 & 99.2 & ~ & 100 \\ 
AlAgS$_2$ (1:1:2) & $P$3$m$1 & 86.2 $\pm$2.0 & 88.9 & 92.4 & 99.7 \\ 
~ & ~ & ~ & ~ & ~ & [98.7] \\ 
TiAl$_3$ (2:6) & $I$4/$mmm$ & 94.7 $\pm$1.2 & 95.5 & ~ & 99.9 \\
Zr$_2$CuSb$_3$ (2:1:3) & $P$–4$m$2 & 84.4 $\pm$12.5 & 97.9 & 99.3 & 100 \\ 
~ & ~ & ~ & ~ & ~ & [99.8] \\ 
Mo$_2$C (8:4) & $Pbcn$ & 96.0 $\pm$1.6 & 98.6 & ~ & 99.9 \\
LaTaO$_4$ (2:2:8) & $Cmc$2$_1$ & 90.0 $\pm$4.7 & 96.6 & 96.7 & 99.9 \\ 
~ & ~ & ~ & ~ & ~ & [99.8] \\ 
ZrO$_2$ (4:8) & $P$2$_1$/$c$ & 97.7 $\pm$0.6 & 98.4 & ~ & 100 \\ 
Li$_5$BiO$_5$ (5:1:5) & $Cm$ & 73.6 $\pm$4.5 & 79.1 & 92.0 & 99.4 \\ 
~ & ~ & ~ & ~ & ~ & [83.0] \\ 
\botrule
\end{tabular*}%
\footnotetext[a]{ The values within parentheses () indicate the number of atoms that participated in the crystal structure creation.}
\footnotetext[b]{ Crystal morphing was performed for some systems which were considered not achieving a sufficiently high $S_\textrm{cos}$ by evolutionary algorithms. The crystal morphing was performed when the mean value of $S_\textrm{cos}$ was $\leq$ 90\% after rounding off the decimal point.}
\footnotetext[c] { BBO-Rietveld and symmetrization.}
\end{minipage}
\end{center}
\end{table}

\subsection*{Evolutionary algorithm for optimizing similarity of XRD pattern to target}\label{subsec7}
First, the evolutionary algorithm's performance was tested whether it could create the crystal structure possessing the target XRD pattern. 
Herein, the target XRD patterns are simulated from crystal structures in ICSD, and the result of the same test for the powder XRD will be shown later.
As the input, the number of atoms was set to be the same as that of the primitive or conventional cell for the target structure. 
Each evolutionary algorithm was performed separately five times. 
The crystal structure with the highest $S_{\textrm{cos}}$ is the best structure to reproduce the target XRD pattern. 
Note that considering the thermodynamic stability, which information was obtained from the first-principles calculations, structures with a formation energy of 0.2 eV/atom higher than the most stable structure in each trial were excluded despite exhibiting the highest $S_{\textrm{cos}}$.

Table \ref{tab:main} shows the result of $S_{\textrm{cos}}$ of the created crystal structure by the evolutionary algorithm and other post process (will be discussed later). 
In many cases, the evolutionary algorithm successfully produced structures with a significantly high $S_{\textrm{cos}}$. 
Among twelve systems, ten exhibited the best structure with the highest $S_{\textrm{cos}}$ $\geq$ 95\% from all five trials of the evolutionary algorithm. 
Six systems exhibited the best structure with mean $S_{\textrm{cos}}$ $\geq$ 95\%. 
Therefore, the evolutionary algorithm is a strong tool for directly creating the crystal structure reproducing the XRD pattern.

\begin{figure*}[b]
 \begin{center}
  \includegraphics[width=0.95\linewidth]{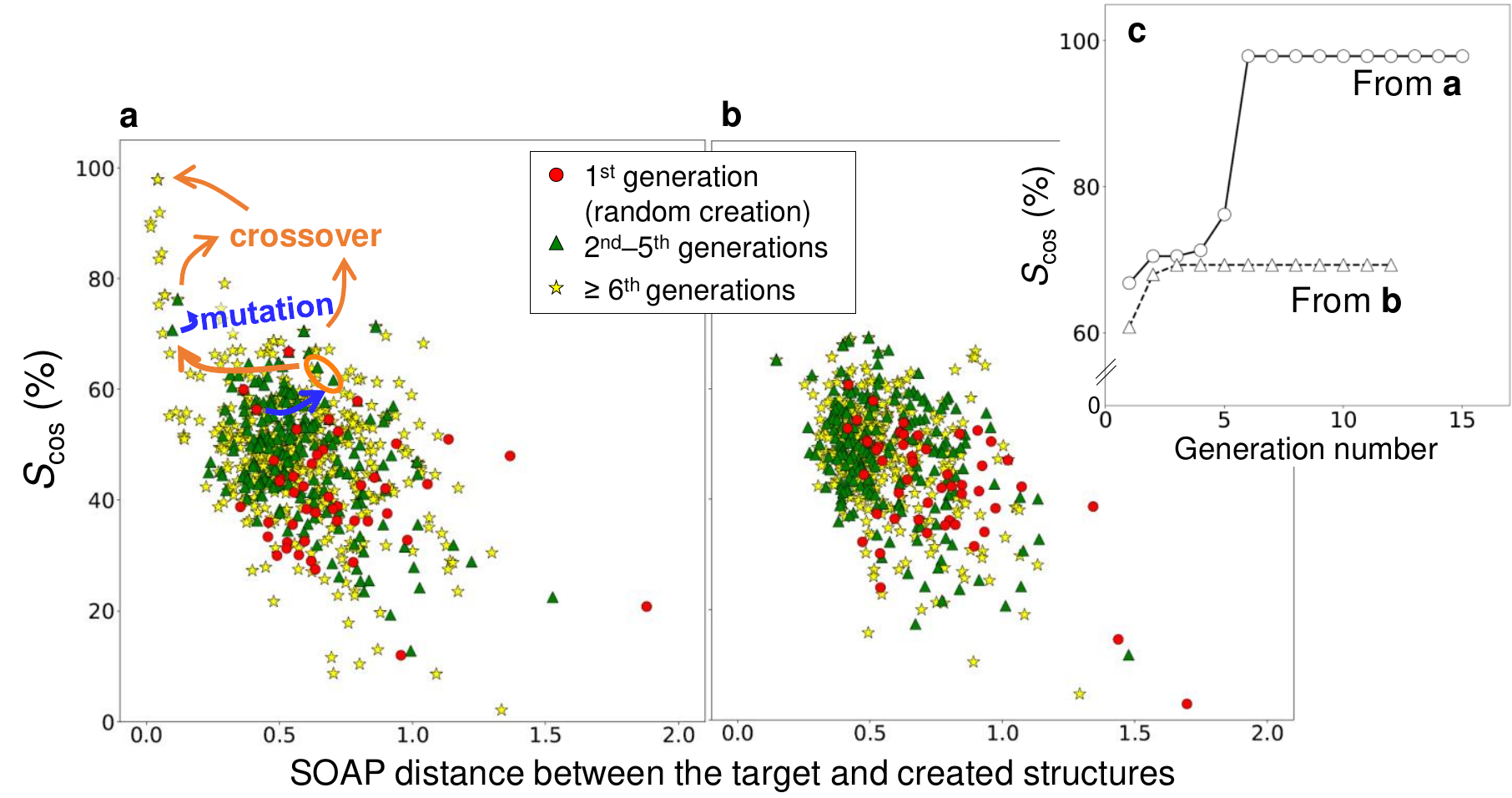}
 \caption{
  \textbf{Different creations of crystal structures for the same target system by an evolutionary algorithm.}
  \textbf{a} Successful and \textbf{b} failed examples of crystal structure creations for achieving high $S_\textrm{cos}$ by the evolutionary algorithm for Zr$_2$CuSb$_3$ system. 
  For the champion structure with $S_\textrm{cos}$ of 98\%, blue and orange indicate the trajectory of the structural evolution by mutation and crossover, respectively. 
  \textbf{c} Increase in $S_\textrm{cos}$ with the increase in generation number for the case of \textbf{a} and \textbf{b}. 
  The maximum $S_\textrm{cos}$ for each generation was connected by lines.
 }
\label{fig:different-evolution}
\end{center}
\end{figure*}

One feature of this algorithm is that it relies on luck in selecting evolutionary paths. 
For example, for Zr$_2$CuSb$_3$ and Zr$_3$Cu$_4$Si$_2$ systems, 
the standard deviation value of $S_\textrm{cos}$ for the five different evolutionary algorithms was $\geq$ 6\%. 
In an evolutionary algorithm, the "luck" is related to the random selection factors that influence the creation of a crystal structure, such as the randomly chosen space group, as well as determining which parent structures will undergo which genetic operators.
If it is possible to specify the exploration path close to the correct answer, such as crystal system and space group, the dependence on the luck can be reduced.
Figure \ref{fig:different-evolution} shows different $S_\textrm{cos}$ distribution of all created Zr$_2$CuSb$_3$ crystal structures during two different trials of the evolutionary algorithm. 
To quantify the structural difference between the created structures and target structure, 
a smooth overlap of atomic position (SOAP)\cite{SOAP1,SOAP2} distances $d$ (see the definition at \nameref{secmethod} section and Oba and Kajita\cite{morphing}) are plotted together.
Herein, a simple concept for the SOAP distance is essential: a smaller distance indicates a slight difference between two structures. 
In addition, note that, if a target structure is unknown (only the target XRD pattern exists), the SOAP distance between the created and the target structure is indefinable.
Therfore, herein, the SOAP distance, defined from the target structure, was used to clearly indicate that the structure with the high $S_\textrm{cos}$ is successfully created close to the target for the purpose of this method examination.

For the successful case, some structures with increased $S_\textrm{cos}$ were proposed by meaningful genetic operators despite randomly created structures having low $S_\textrm{cos}$ at the first generation. 
In the result, as shown in Fig. \ref{fig:different-evolution}c, $S_\textrm{cos}$ was gradually increased with increasing a generation number.
Finally, a structure with a significantly high $S_\textrm{cos}$ of 98\% was created. 
However, for the unfortunate case, a high $S_\textrm{cos}$ was not obtained because a meaningful structural evolution was not performed and remained at a similar level ($\leq$ 69\%) with those from structures created at early generation. 
Therefore, to increase the probability of successfully creating the structure that matches the desired goal, it is recommended to make multiple attempts using the evolutionary algorithm and complement any unsuccessful trials. 
\subsection*{Crystal morphing with Bayesian optimization for optimizing similarity of XRD pattern to target}\label{subsec8}
Despite multiple tries of the evolutionary algorithm, it could fail to create crystal structure to reproduce the target XRD pattern for some complex systems. 
Particularly, the mean and maximum $S_\textrm{cos}$ remained only 74\% and 79\% for Li$_5$BiO$_5$ system, respectively. 
Crystal morphing with Bayesian optimization was applied to such five systems with a mean value of $S_\textrm{cos}$ $\leq$ 90\% after rounding off the decimal point (see comment in Table \ref{tab:main}) from five different evolutionary algorithms. 
For other systems with a high mean value of $S_\textrm{cos}$, the created structures used as the input structures for crystal morphing are already close to the target structure and similar to each other. 
Therefore, the interpolation between such similar structures is not considered meaningful.

\begin{figure*}[htp]
 \begin{center}
  \includegraphics[width=0.95\linewidth]{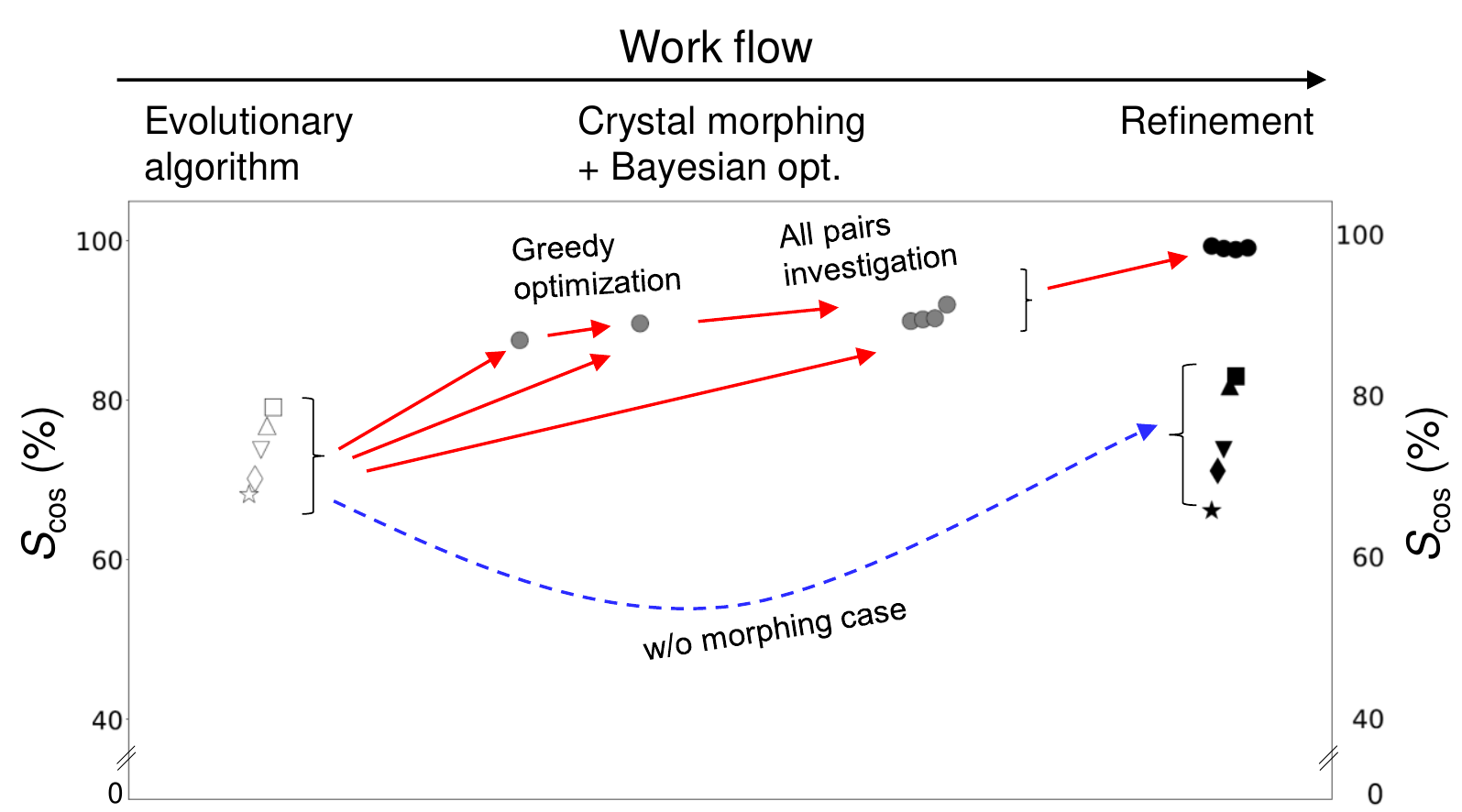}
 \caption{
  \textbf{Sequential procedure of increasing $S_\textrm{cos}$ for Li$_5$BiO$_5$ system.} 
  Red solid and blue dashed indicators show the procedures with and without the crystal morphing with Bayesian optimization, respectively. 
  Refinement (including procedures both BBO-Rietveld and symmetrization) could not increase $S_\textrm{cos}$ for the case where $S_\textrm{cos}$ of input structures provided by evolutionary algorithm remained only $\sim$80\%. 
  The refinement succeeded by improving it to $>$ 90\% by crystal morphing with Bayesian optimization, and $S_\textrm{cos}$ reached $>$ 99\%. 
  Different marks for evolutionary algorithm indicate $S_\textrm{cos}$ obtained from different trials. 
  For crystal morphing and post Refinement, structures with $S_\textrm{cos}$ higher than those of previous methods are shown among the obtained.
 }
\label{fig:Li5BiO6-success}
\end{center}
\end{figure*}

The result of the five systems shows that crystal morphing with Bayesian optimization proposed crystal structures with further increased $S_\textrm{cos}$ than those of the input structures (see Table \ref{tab:main}). 
In particular, Li$_5$BiO$_5$ system achieved $S_\textrm{cos}$ of 89\% and 92\%, which were significantly increased from 79\%, 
by greedy optimization and followed all pairs investigation, respectively. 
This procedure is shown in Fig. \ref{fig:Li5BiO6-success}. 
The improvement of $S_\textrm{cos}$ by crystal morphing means that it is a proper complementary method to explore space that the evolutionary algorithm could not. 
However, the search space by this method relies on selecting the input structures. 
Therefore, it is significant to select multiple input structures having sufficiently high $S_\textrm{cos}$ but are different from each other for exploring vast space.

\subsection*{Refinement}\label{subsecrefine}
As the last step, refinement was performed to decrease further the gap between XRD patterns of the created structure and target. 
For a successful Rietveld refinement, preparing an initial structure that has a slight difference in the target XRD patterns, namely, a high $S_\textrm{cos}$, is essential. 
Except for Li$_5$BiO$_5$, for the eleven systems with the maximum $S_\textrm{cos}$ $\geq$ 88\% obtained by evolutionary algorithm, Rietveld refinement raised $S_\textrm{cos}$ to $\geq$ 97\%. 
In addition, for Li$_5$BiO$_5$, when the structure with $S_\textrm{cos}$ $\geq$ 92\% obtained by crystal morphing was used as input of Rietveld refinement, $S_\textrm{cos}$ increased to 99\%. 
Notably, the performance of Rietveld refinement without crystal morphing did not increase or even decrease $S_\textrm{cos}$ when poor input structures with $S_\textrm{cos}$ $<$ 80\% were prepared by evolutionary algorithm, as shown in Fig. \ref{fig:Li5BiO6-success}. 
Therefore, creating search space and increasing $S_\textrm{cos}$ by crystal morphing is helpful for a successful post–refinement process in the case that the input structure could not achieve sufficiently high $S_\textrm{cos}$. 
 
Then, symmetrization was performed after Rietveld refinement. 
It further increases $S_\textrm{cos}$ of the refined structure during determining its space group and tuning the structure with high symmetry. 
For all the twelve systems, finally, our scheme achieved significantly high $S_\textrm{cos}$ $>$ 99\%. 
Therefore, it is concluded that the crystal structure that reproduces the XRD pattern is successfully created. 

\subsection*{The test by the experimentally-measured target XRD}\label{subsecexp}
Until now, the target XRD patterns were simulated using crystal structures included in ICSD.
In actual case that the target XRD pattern is a measured one after synthesis, it includes other effects such as background noises and instrumental artifacts.
Therefore, for an additional test, four powder XRD patterns were employed as the target.

Table \ref{tab:vs-exp} shows the result of $S_\textrm{cos}$ of the created crystal structure.
Compared with the case where the target XRD pattern was simulated from the structure in ICSD, the $S_\textrm{cos}$ reached to only 80\%--97\%. 
The degraded performance is mainly ascribed to the existence of background. 
Because the background hindered a creation of structure to reproduce the target XRD pattern correctly, the same test was performed after removing the background.
In the result, the $S_\textrm{cos}$ reached to greater than 96\%. 
The space group of the created structures were the same as the correct one.
If the background is removed, the XRD pattern becomes almost equivalent to the simulated one which has clearer peaks; therefore, the suggested algorithm can work better.


\begin{table}[htp]
\begin{center}
\begin{minipage}{\textwidth}
\caption{
$S_\textrm{cos}$ obtained by crystal structure creation methods for four raw XRD patterns without and with background noise. The values after $\pm$ indicate the standard deviation of $S_\textrm{cos}$ from five different trials of the evolutionary algorithm.
}\label{tab:vs-exp}

\begin{tabular*}{\textwidth}
{@{\extracolsep{\fill}}lccccccc@{\extracolsep{\fill}}}
\toprule%
\multicolumn{3}{@{}c@{}}{Target} & \multicolumn{4}{@{}c@{}}{Highest $S_\textrm{cos}$ of XRD pattern}\\
\multicolumn{3}{@{}c@{}}{~} & \multicolumn{4}{@{}c@{}}{ of created structure with respect to target (\%)}\\
\cmidrule{1-3}\cmidrule{4-7}
Materials \footnotemark[1] & Space & Background & \multicolumn{2}{@{}c@{}}{By evolutionary}  & By Crystal & By Refine- \\
~ & group & noise & \multicolumn{2}{@{}c@{}}{}{algorithm} & Morphing & ment \footnotemark[2] \\
\cmidrule{4-5}
~ & ~ & ~ & Mean & Max. & ~ &\\
\midrule%
ZnAl$_2$O$_4$  & $Fd$–3$m$ & raw & 79.4 $\pm$0.2 & 79.5 & 79.7 & 79.7 \\
(2:4:8) & ~ & removed & 98.7 $\pm$0.2 & 98.9 & 98.9 & 98.9 \\
\midrule%
Al$_2$O$_3$  & $R$–3$c$ & raw& 96.2 $\pm$0.3 & 96.3 & 96.6 & 96.6 \\
(4:6) & ~ & removed & 98.4 $\pm$0.3 & 98.6 & 98.8 & 98.8 \\
\midrule%
CaTiO$_3$  & $Pnma$ & raw & 87.9 $\pm$0.3 & 88.2 & 88.4 & 88.4 \\
(4:4:12) & ~ & removed & 97.4 $\pm$0.5 & 97.9 & 97.9 & 97.9 \\
\midrule%
ZrO$_2$  & $P$2$_1$/$c$ & raw & 92.6 $\pm$0.3 & 93.0 & 93.0 & 93.1 \\ 
(4:8) & ~ & removed & 95.1 $\pm$0.8 & 96.3 & 96.4 & 96.4 \\
\botrule
\end{tabular*}%
\footnotetext[a]{ The values within parentheses () indicate the number of atoms that participated in the crystal structure creation.}
\footnotetext[b] { BBO-Rietveld and symmetrization.}
\end{minipage}
\end{center}
\end{table}

\section*{DISCUSSION}\label{secdiscussion}
The target and created XRD pattern and their structures are summarized in Supplementary Figs. \ref{fig:cifxrd1} and \ref{fig:cifxrdexp1}.  
It is notable that for two systems, despite significantly high $S_\textrm{cos}$ $>$ 99\%, created crystal structures had distinct parts from the target ones. 
Zr$_3$Cu$_4$Si$_2$ had an exchange of Cu and Si sites, and Li$_5$BiO$_5$ had shifted O layers. 
This is a limitation of conventional XRD analysis to distinguish such structures, and support by other analyses, which focus on the local structure and nearest neighbors such as X-ray absorption fine structure\cite{newville2001exafs,zheng2018automated}, might be useful to determine the structure more correctly.

The performance of Evolv\&Morph is compared with that obtained only by the Rietveld refinement after DB-search. 
To prepare the vanilla strategy for comparison, first we selected close structures to the targets from Materials Project Database (MPD)\cite{MPD} and SpringerMaterials (SM)\cite{springer-site} database.
Then, BBO-Rietveld and symmetrization were performed after the lattice volumes with the highest $S_\textrm{cos}$ were adjusted. 
The lattice parameters and internal coordinates of atoms of a structure loaded from other DB have only a small difference from those of the target one. 
Note that MPD and SM record the structures obtained by first-principles calculations and experiments, respectively.
Namely, the vanila strategy was advantageous because it could start from the input structures significantly close to the target.
Nevertheless, Evolv\&Morph successfully exhibited $S_\textrm{cos}$ similar to or higher than those obtained by the vanilla strategy. 
This result also indicates that the Evolv\&Morph may produce structures well matched with the target XRD even from scratch. 
More details are discussed in Supplementary section \nameref{subsec9}.

The suggested algorithm can be performed in an ordinary personal computer (PC), except for the first-principles calculations which are commonly performed in parallel on other cluster computers for high performance computing system. 
Evolutionary algorithm was simply performed on a single core of an ordinary PC (Intel\textsuperscript{\textregistered} Xeon\textsuperscript{\textregistered} Gold 6248 central processing unit (CPU) @3.00GHz with memory of 32 GBytes).
Multiple attempts of evolutionary algorithm could be performed in multi-core processors in the current PC separately.
Crystal morphing, which can be performed on a single core, was performed with four cores in parallel with Bayesian optimization for fast sampling of proposed structures in the same PC.
On the other hand, first-principles calculations were performed in a multi-core cluster computing system (AMD EPYC\textsuperscript{\texttrademark} 7742 CPU @2.25GHz), where multiple calculations can be executed with eight cores in parallel.
Each attempt of evolutionary algorithm took 0.5--1.5 days, including the optimizations of structures by the first-principles calculations.
The crystal morphing for the all-pairs investigation took only a few hours.
Therefore, the suggested algorithm does not require significant costs for searching a usual unit cell.

As a supporting structural optimizer for each structure created by evolutionary algorithm, first-principles calculations were employed. 
It helps to optimize the created structure to become more stable structure with removing many non-physical bonds.
In addition, many small XRD peaks from unstable and less symmetric structures can be removed if the structure is relaxed to more symmetric structures. 
It also provides thermodynamic stability information to consider which created structure is more realistic when multiple structures exist with high and similar $S_\textrm{cos}$.
However, the usage of the first-principle calculations brings some demerits. 
One is that its structural optimization based on the thermodynamics does not guarantee an increase in the similarity score of the XRD pattern for the target.
Sometimes, the structural optimization might decrease $S_\textrm{cos}$ in the opposite direction to our goal. 
Another one is the requirement of additional time cost. 
For the evolutionary algorithm, the time cost incurred during the first-principles calculations is significantly greater than the time cost incurred without the first-principles calculations.
Therefore, the usage of the first-principles calculations makes a limit to apply the present algorithm to much larger systems.
If much faster structural optimization methods such as machine learning interatomic potential\cite{ontheflymlp,m3gnet} become satisfactorily robust and accurate, they will be able to replace the role of the supporting structural optimizer. 

In this study, an automated crystal structure creation method, consisting of the evolutionary algorithm and crystal morphing supported by Bayesian optimization (Evolv\&Morph), was proposed for reproducing the XRD pattern.
The method optimizes the similarity score of XRD patterns of the created structures and the target.
The evolutionary algorithm can automatically create various structures by genetic operators without input structures.
Crystal morphing, using the input structures obtained by the evolutionary algorithm, 
expands search space and further increases the similarity of XRD patterns creating a better input structure for the post–refinement.
After the refinement, for twelve binary and ternary systems in different crystal structures where the target XRD patterns were provided by simulation from the measured structures, Evolv\&Morph achieves cosine similarity of $\sim$99\%.
In addition, for four different powder XRD patterns, the present method achieved cosine similarity of of $\sim$96\%, after removal of background.
These results indicate that the created crystal structure successfully reproduces the target XRD pattern.
Therefore, the present method can identify unknown crystal structures after getting XRD measurement without depending on DB.
Furthermore, Evolv\&Morph can play a role of inverse design,\cite{inversedesign,inversedesign2} which indicates that the desired property is defined first and the materials with such property are automatically searched.
The optimization target score, which corresponds to the cosine similarity of the XRD pattern used in this study (see Fig. \ref{fig:overview}), can be exchanged for particular functional property according to the goal of the material design; 
therefore, the present method has a powerful potential to be applicable to material design as well as crystal structure determination.

\section*{METHODS}\label{secmethod}
\subsection*{Evolutionary algorithm}\label{subsec2}
\begin{figure*}[htp]
 \begin{center}
  \includegraphics[width=0.95\linewidth]{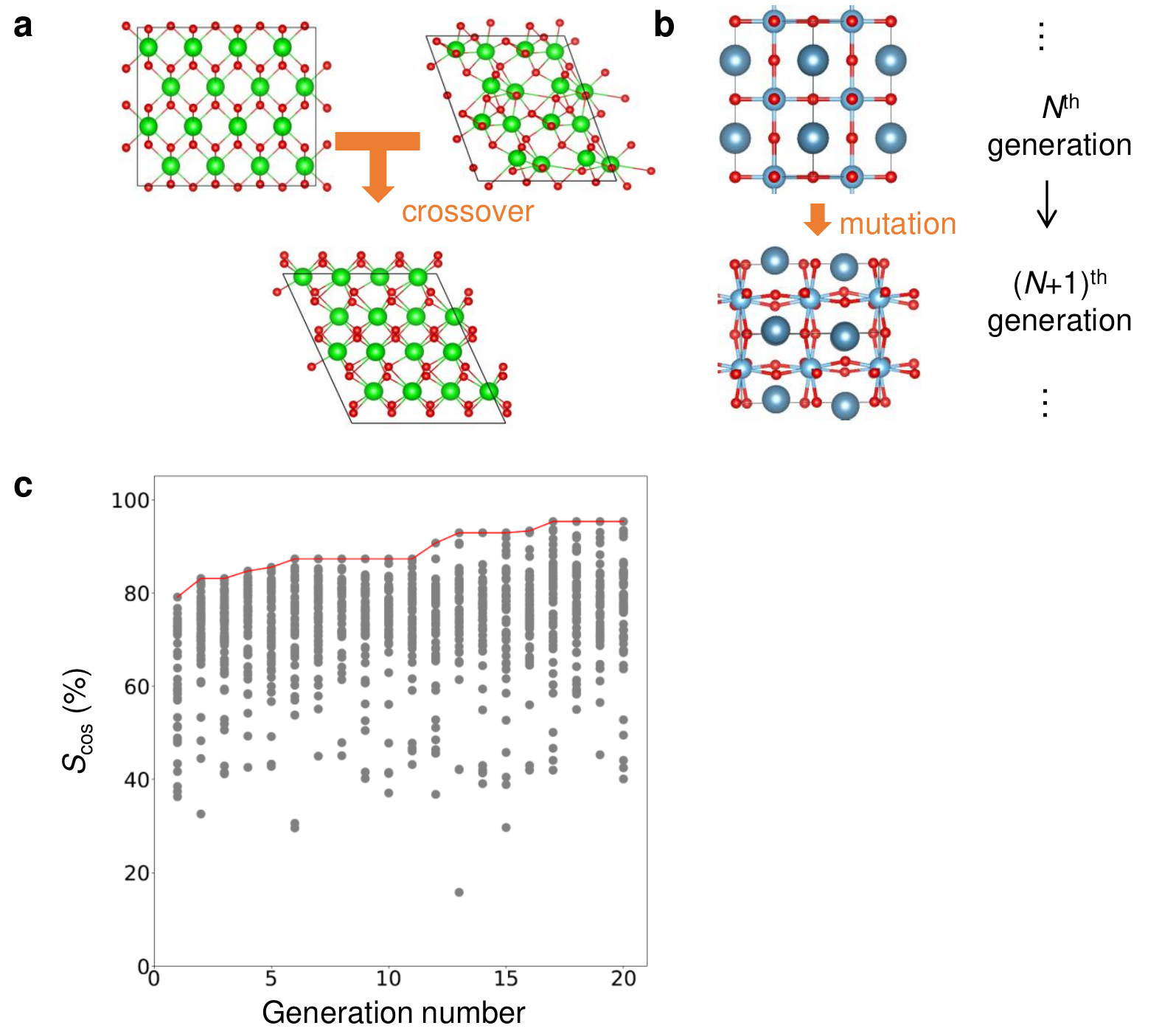}
 \caption{
  \textbf{Evolutionary algorithm.} Example of crystal structure creations by genetic operator \textbf{a} crossover and \textbf{b} mutation in an evolutionary algorithm. 
  This way, the structures in (\textit{N}+1)-th generation can be derived from those in $N$-th generation. 
  \textbf{c} Example of optimization of $S_\textrm{cos}$ according to increase of generation number. 
  Red lines connect each generation's highest $S_\textrm{cos}$.
 }
\label{fig:evolutionary}
\end{center}
\end{figure*}

An evolutionary algorithm is one of the optimization methods inspired by biological evolution such as reproduction, mutation, recombination and selection \cite{USPEX1}. 
It automatically creates various crystal structures by genetic operators such as crossover (shown in Fig. \ref{fig:evolutionary}a) and mutation (shown in Fig. \ref{fig:evolutionary}b) for multiple generations (ages). 
Crossover, also usually called heredity or two-parent variation operator, creates an offspring structure by combining spatially coherent slabs in terms of fractional coordinates of two parent structures with averaging two parent lattice vector matrices. 
Mutation creates another structure by applying variations on one parent structure, such as lattice distortion and exchange of atoms.

In addition, the evolutionary algorithm includes an optimization function because it repeats the following procedure for multiple generations: 
it remains some survivors with high scores, kill some losers with low scores in a generation, and creates additional structures in the next generation. 
As shown in Fig. \ref{fig:evolutionary}c, by setting the target property (also known as a fitness) to \textit{S}$_\textrm{cos}$, this algorithm tries to create various crystal structures to possess XRD patterns more similar to the target one. 
Input structures are not essential for this method, and structures can be searched in a vast space. 
Moreover, thermodynamic calculations such as first-principles calculations can aid the evolutionary algorithm by allowing optimization of the created crystal structure onto a local minimum energy surface, 
thereby preventing the formation of unphysical structures.

USPEX (Universal Structure Predictor: Evolutionary Xtallography) \cite{USPEX1,USPEX2,USPEX3,USPEX-code} code was used for the evolutionary algorithm program. 
In the first generation, crystal structures were produced using randomly selected space groups. 
When the structures were randomly created, the minimum bond lengths of 1.95 \AA\ and 1.5 \AA\ were limited between the same and different elements, respectively. 
The space group numbers were limited to 3–230.
From the second-generation onwards, new crystal structures were produced by genetic operators: crossover (50\%), random symmetry creation (20\%), and mutation (30\%). 
As mutation operators, permutation (10\%) and softmutation (20\%)\cite{softmutation} were used, where the former exchanges the occupied sites of atoms, and the latter moves atoms along to the eigenvector of the softest modes.

Each generation consisted of fifty crystal structures. 
The evolutionary algorithm was terminated if the best-ranked crystal structure was not changed over ten generations or the generation number reached 20th.

\subsection*{First-principles calculations}\label{subsec11}
First-principles calculations were performed for each crystal structure to be optimized after the evolutionary algorithm created it to reduce the chances of becoming unphysical or too unstable structures.
All first-principles calculations were performed using the projector augmented wave (PAW) \cite{PAW1,PAW2} method implemented in the Vienna Ab initio Simulation Package (VASP) \cite{VASP1,VASP2}. 
We used the exchange-correlation function of the generalized gradient approximation (GGA) parameterized in the Perdew–Burke–Ernzerhof (PBEsol) form modified for solids \cite{PBEsol}. 
Focusing on high efficiency, a low cutoff energy of 300 eV and a low $k$-space resolution of 0.12 (in a unit of $2\pi$/\AA) were used. 
Optimizing the created unit cell was performed until the interatomic force on each atom was reduced to within 0.03 eV/\AA\ or the number of ionic iterations reached 30. 
The wall time for stopping the unfinished calculation was also tightly set to ten minutes with eight cores in parallel, 
to avoid frequent time losses caused by convergence failures that often occur in calculations for unphysical structures.


\subsection*{Crystal morphing with Bayesian optimization}\label{subsec3}
\begin{figure*}[htp]
 \begin{center}
  \includegraphics[width=0.74\linewidth]{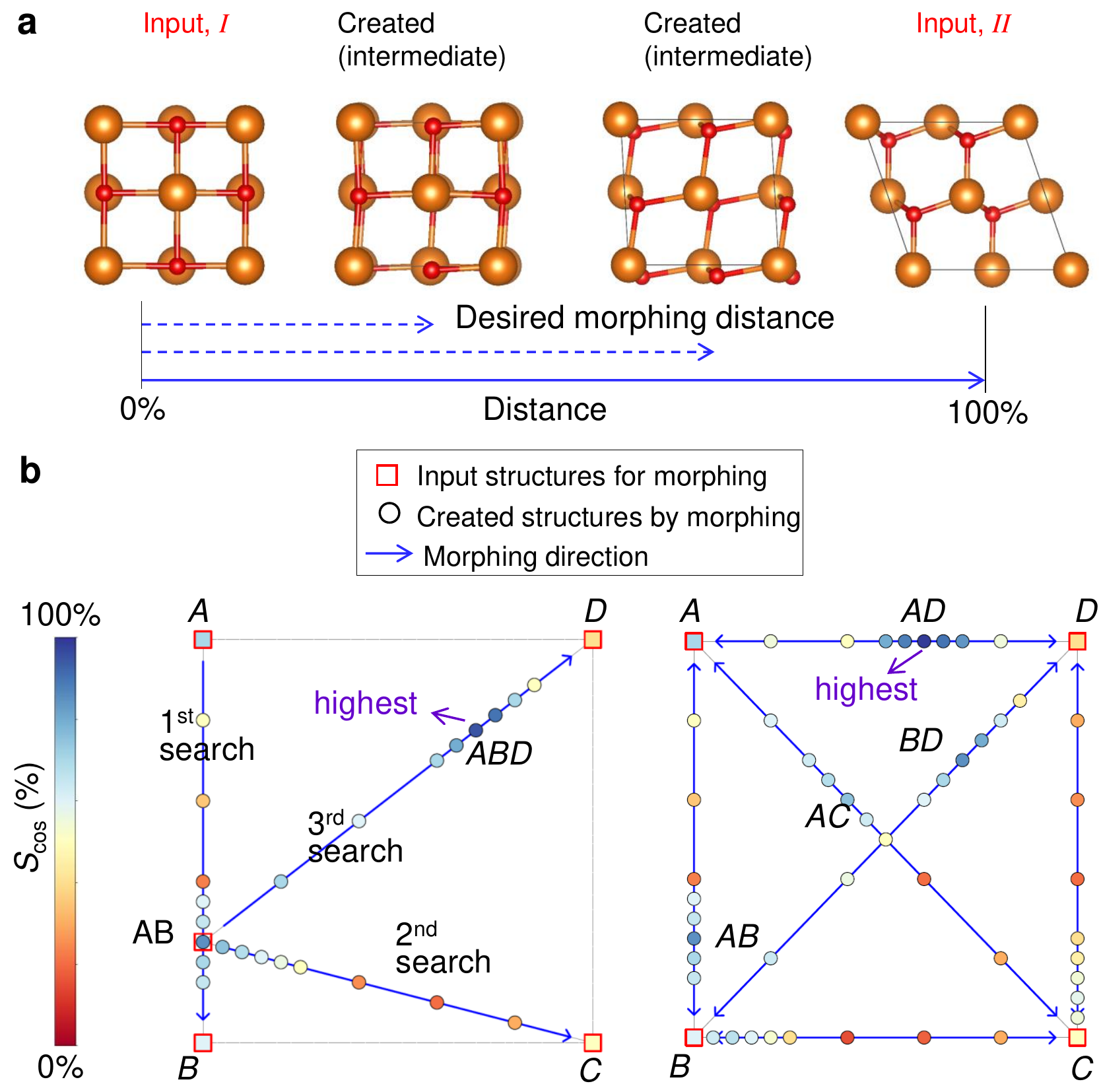}
 \caption{
  \textbf{Crystal morphing.} 
  \textbf{a} Concept of crystal morphing.
  By morphing structure $I$ towards structure $II$, it is capable of creating intermediate crystal structures on the desired morphing distance. 
  A distance can be determined based on structural descriptor between two input crystal structures.
  The desired morphing distance can be suggested by Bayesian optimization which searches crystal structures with high $S_\textrm{cos}$.
  Then, crystal morphing can create intermediate structures possessing the structural descriptor on the desired distances.
  In the process, the lattice vectors and internal coordinates of atoms of the intermediate structures on the desired morphing distance are manipulated.
  \textbf{b} A search of crystal structures possessing higher \textit{S}$_\textrm{cos}$ by crystal morphing and Bayesian optimization via (left panel) “greedy optimization” and (right panel) “all pairs investigation”.
  The detailed procedure is described in the main text.
  A red rectangular mark indicates the input structures for crystal morphing.
  A circle mark indicates an intermediate structure found by crystal morphing and Bayesian optimization. 
  The color indicates the \textit{S}$_\textrm{cos}$ of the structure.
  In this caption, an example case is explained.
  (left panel) greedy optimization: \textit{A}, \textit{B}, \textit{C}, and \textit{D} are the input structures.
  Their \textit{S}$_\textrm{cos}$ are assumed to be \textit{A} $>$ \textit{B} $>$ \textit{C} $>$ \textit{D}.
  The first search by crystal morphing and Bayesian optimization is performed between \textit{A} and \textit{B}.
  Intermediate structure \textit{AB} is newly found with higher \textit{S}$_\textrm{cos}$ than those of \textit{A} and \textit{B}.
  Then, a second search is performed between \textit{AB} and \textit{C}.
  The intermediate structure with higher \textit{S}$_\textrm{cos}$ than those of \textit{AB} and \textit{C} is not found.
  Then, a third search is performed between \textit{AB} and \textit{D}. The intermediate structure \textit{ABD} with higher \textit{S}$_\textrm{cos}$ than those of \textit{AB} and \textit{C} is newly found.
  (right panel) All pairs investigation: \textit{A}, \textit{B}, \textit{C}, and \textit{D} are the input structures.
  The search by crystal morphing and Bayesian optimization is performed among the combinations of each pair.
  The intermediate structures, \textit{AB}, \textit{AC}, \textit{AD}, and \textit{BD} with higher \textit{S}$_\textrm{cos}$ than those of each pair of two input structures are newly found. 
 }
\label{fig:morphing}
\end{center}
\end{figure*}

Crystal morphing \cite{morphing} is an interpolation method to create intermediate crystal structures between two input structures ($I$ and $II$) shown in Fig. \ref{fig:morphing}a.
By morphing structure $I$ towards structure $II$, it is capable of  creating intermediate crystal structures on the desired morphing distance determined by crystal structure descriptors such as SOAP \cite{SOAP1,SOAP2}.
Lattice vectors and internal coordinates of atoms of the intermediate structures on the desired morphing distance are manipulated by interpolation.
The metric, called SOAP distance, promises that the morphing takes into account invariances in translation, rotation, and unit-cell choice.

Bayesian optimization is a sequential design strategy to find the maximum or minimum value of targeting property based on a posterior distribution and an acquisition function \cite{bayesian}.
Therefore, when \textit{S}$_\textrm{cos}$ is employed as the optimization target, this optimization method searches an optimal \textit{S}$_\textrm{cos}$, which corresponds to the desired morphing distance, in the line drawn by the crystal morphing. 
Although crystal morphing is an interpolation method, search space can be expanded by multiple trials by selecting different input structures. 
We used two recipes to find a structure with high \textit{S}$_\textrm{cos}$, as shown in Fig. \ref{fig:morphing}b. 

First one is “greedy optimization”.
The input structure list is prepared, and the structures are sorted in descending order for their \textit{S}$_\textrm{cos}$ scores. 
The two structures with the highest \textit{S}$_\textrm{cos}$, as indicated by “$A$” and “$B$” in Fig. \ref{fig:morphing}b, are selected as for the first search. 
If an intermediate structure with the \textit{S}$_\textrm{cos}$ is higher than the two input structures, the found one is added to the input structure list, but the input two structures are removed from the list. 
If such an intermediate structure is not found, the input structure with the second highest \textit{S}$_\textrm{cos}$ is excluded from the input structure list.
Then, the two structures with the highest \textit{S}$_\textrm{cos}$ are reselected for the next search. 
This procedure is repeated, and the structure with higher \textit{S}$_\textrm{cos}$ is sequentially updated.
This method works \textit{S}$_\textrm{cos}$ score maximization successfully if higher \textit{S}$_\textrm{cos}$ are gradually updated by newly found intermediate structures. 
However, suppose the intermediate structures have lower \textit{S}$_\textrm{cos}$ than the input structures.
In that case, the method fails to optimize \textit{S}$_\textrm{cos}$ by searching the limited space as a consequence.
To avoid such limited search space, the second method, “all pairs investigation” was used.
This method searches the intermediate structures among all pairs formed in the input structure list. 
As a result, it can expand search space.

Greedy optimization was first used with input structures obtained by the \textit{S}$_\textrm{cos}$ champions in five different trials of evolutionary algorithms, 
and then followed by all pairs investigation. 
Newly created intermediate structures by greedy optimization with higher \textit{S}$_\textrm{cos}$ than the input structures were also added to the input structure list for all pairs investigation with the structures obtained by evolutionary algorithms. 
For the input structure selections and expanded search, one can repeat the mentioned two methods with increased structures; however, in this study, only one cycle was used.

Squared SOAP distance $d$ between two structures $i$ and $ii$ can be defined by a following equation:
\begin{eqnarray}
d^2(\chi_i, \chi_{ii}) \equiv
\frac{\lvert \mathbf{P}(\chi_i) - \mathbf{P}(\chi_{ii})\rvert^2}
{\lvert \mathbf{P}(\chi_i)\rvert \lvert \mathbf{P}(\chi_{ii})\rvert},
\label{eq:dps}
\end{eqnarray}
where the vector $\mathbf{P}$ is the SOAP power spectrum
and $\chi$ is the reciprocal-space representation of the structure $i$ or $ii$.
For SOAP parameters, the width of the Gaussian
function $\sigma = 0.5$~\AA~was used. 
The maximum sizes of the radial basis functions and spherical harmonics were 10 and 6, respectively.
For morphing an input structure $I$ towards another input structure $II$ to create an intermediate structure located at the desired SOAP distance, the optimization of reciprocal lattices and internal coordinates of atoms was performed by the steepest descent method with updating their step sizes of 0.02.
The maximum iteration number was set to fifteen.
The type of elements was distinguished by taking a different sign for the elements of the real-space density distribution for the two-element case. 
We decomposed the system into three pairwise systems for the three-element case and defined a distance as a sum of each pairwise system.
Other parameters and more detailed theory relevant to crystal morphing can be referred to Oba and Kajita. \cite{morphing} 

Gaussian processes framework in python optimization (GPyOpt) \cite{GPyOpt} code was used for Bayesian optimization.
The interpolated SOAP distance point between two input structures at 0, 25, 50, 75, 100\%, and two randomly selected points were investigated to get an initial posterior distribution. 
Then, four additional iterations were performed with each four-point parallel investigation.

\subsection*{Refinement}\label{subsec13}

The created structure can be further tuned by refinement method such as Rietveld refinement \cite{bayesian} and symmetrization. 
Rietveld refinement tunes the structure to decrease the gap between simulated and target XRD patterns. 
Therefore, it can raise the $S_{\textrm{cos}}$ score.
Symmetrization is refining a crystal structure to have symmetry when determining a space group. 
Ideally, a crystal structure has a unique space group; however, the determined space group could be slightly changed according to a tolerance parameter to satisfy all given crystallographical constraints. 
Therefore, technically, multiple refined crystal structures can be obtained according to changing the tolerance parameter. 

BBO-Rietveld code \cite{BBO-rietveld} was used for Rietveld refinement.
Because Rietveld needs a lot of parameters, the refinement result strongly depends on the setting of parameters. 
BBO-Rietveld tests various combinations of parameters and suggests the best refined crystal structure determined as the case where the weighted profile residual factor is the lowest. 
The iteration number for optimization was set at 200.
The intermediate structures, which were obtained by crystal morphing, with \textit{S}$_\textrm{cos}$ score higher than input structures were all refined by BBO-Rietveld.

The symmetrization was performed using SPGLIB \cite{spglib} in PHONOPY code \cite{phonopy}. 
The tolerance factor considered is summarized in Supplementary Table \ref{tab:parameters}. 
If multiple space groups were found according to different tolerance factors, the space group with higher \textit{S}$_\textrm{cos}$ score was determined for the refined crystal structure. 
However, if their difference in \textit{S}$_\textrm{cos}$ score is less than 1\%, a space group with higher symmetry was selected.

\subsection*{XRD simulations}\label{subsec14}
The XRD patterns scanned by $\theta$–2$\theta$ mode were simulated using GSAS-II code \cite{GSAS} with a source energy of Copper K–$\alpha$ (wavelength of ~1.54 \AA) between $2\theta$ range of 0–180° with a width of 0.01°.
When a powder XRD pattern was used as the target, the same $2\theta$ range of 5–90° was used for the XRD simulation of the created structures.
Background was removed by subtraction from the powder XRD pattern by using auto peak search function in GSAS-II and distinguishing the parts of peaks and background noise. Negative intensities were changed into positive ones.

\clearpage

\backmatter

\bmhead{CODE AVAILABILITY}
The modified contents in USPEX code for evolutionary algorithm and crystal morphing code are downloadable at https://github.com/ToyotaCRDL/EvolvMorph. Other codes such as the main part of USPEX, BBO-Rietveld, GSAS-II, GPyOpt, and PHONOPY are downloadable from their own repositories.

\bmhead{ACKNOWLEDGEMENTS}
J.L. would like to thank Enago (https\text{://}www.enago.com) for editing and reviewing this manuscript for English language.
The authors also thank R. Jinnouchi and A. Suzumura in TCRDL for providing primary code for SOAP and discussion on Rietveld, respectively.

\bmhead{AUTHOR CONTRIBUTIONS}
J. L. mainly performed simulations and prepared the manuscript. 
J. O. and S. K. developed crystal morphing code. 
N. O. and S. K. designed the project. 
All authors discussed the results and wrote the manuscript.

\bmhead{COMPETING INTERESTS}
The authors declare no competing interests.

\textbf{Correspondence} \textrm{and requests for materials should be addressed to J. L.}

\clearpage

\clearpage

\bibliography{main.bib}

\clearpage
\section*{Supplementary Information}
\beginsupplement

\vspace{3em}

\subsection*{Cosine similarity with isotropic volume changes}\label{csdemerit}
In this section, the evaluation of “similarity” for XRD patterns is discussed. Hern\'{a}ndez-Rivera \textit{et al}.\cite{similarity-compare} quantified the sensitivity of dozens of similarity functions summarized by Cha \cite{Cha2007} on the change of XRD pattern and peak features such as shift, split, and broadening. 
They concluded that there was no metric to be universal most or least sensitive across all types of different peak features. 

\begin{figure*}[htp]
 \begin{center}
  \includegraphics[width=0.95\linewidth]{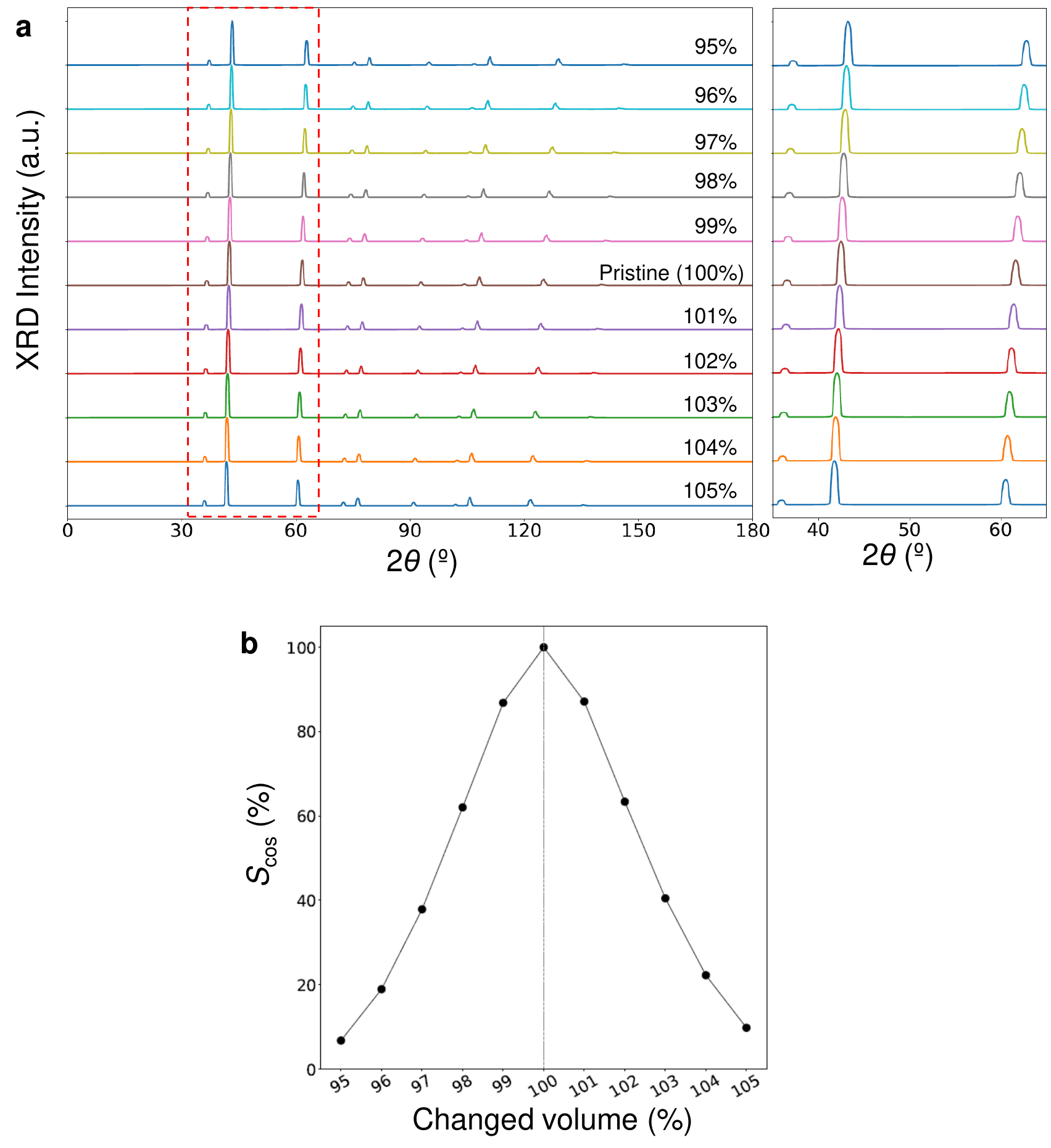}
 \caption{
 \textbf{Sensitivity of $S_\textrm{cos}$ to volume change.}
  Dependence of \textbf{a} XRD pattern and \textbf{b} $S_\textrm{cos}$ (vs. pristine) on volume change of MgO in a Rocksalt structure. 
  At the left panel of \textbf{a}, the number indicates the changed volume. 
  At the right panel of \textbf{a}, XRD patterns at 2$\theta$ of $35$–$65^{\circ}$ (area indicated inside red dashed lines) are magnified. 
  XRD peaks were broadened by Gaussian smearing of $0.5^{\circ}$. 
  As the volume increases isotropically, the XRD peaks are shifted towards the lower angle.
 }
\label{fig:mgoxrd}
\end{center}
\end{figure*}

In this study, cosine similarity (\textit{S}$_\textrm{cos}$) was used as a similarity metric for two XRD patterns. 
However, it has a critical weak point being too sensitive to the peak shift, which corresponds to the change of lattice volume (or parameter) of crystal structure. 
If XRD peaks entirely shift towards higher and lower $2\theta$ regions, it corresponds to decreased and increased lattice parameters, respectively. 
This indicates that $S_\textrm{cos}$ for the two similar structures with only a slight difference in lattice parameter would be low. 
Supplementary Figure \ref{fig:mgoxrd} shows an example of the dependence of XRD pattern and \textit{S}$_\textrm{cos}$ on the isotropic volume change.
$S_\textrm{cos}$ was obtained between the pristine structure and the one with the changed volume.
With the increase in a volume change, \textit{S}$_\textrm{cos}$ significantly decreased. 
\textit{S}$_\textrm{cos}$ became even $<$ 10\% at volume change of ±5\%. 
Therefore, $S_\textrm{cos}$ judges that the pristine structure and the one with only 5\% volume expansion or contraction are completely different.

To this end, we propose an idea to complement the weak point of $S_\textrm{cos}$ for correct evaluation of similarity between XRD patterns by obtaining the maximum value of $S_\textrm{cos}$ after isotropic volume scanning. 
As an example, $\alpha$-Fe$_2$O$_3$ with a space group of \textit{R}–3\textit{c} in a trigonal structure was prepared for generating a target XRD pattern. 
For test samples, $\alpha$-Al$_2$O$_3$ and $\theta$-Al$_2$O$_3$ with space groups of \textit{R}–3\textit{c} and \textit{C}2/\textit{m} in trigonal and monoclinic structures were prepared, respectively. 
The crystal structures are shown in Supplementary Fig. \ref{fig:al2o3xrd}a.
The local structure of $\alpha$-Al$_2$O$_3$ is similar to $\alpha$-Fe$_2$O$_3$ more than $\theta$-Al$_2$O$_3$, 
in terms of the same space group and intuitive appearance. 
The XRD patterns of Fe$_2$O$_3$ were obtained using Al$_2$O$_3$ crystal structures after substitutions of Al for Fe sites. 
As shown in Supplementary Fig. \ref{fig:al2o3xrd}b, the number and peak patterns in XRD of Fe$_2$O$_3$ in the $\alpha$-Al$_2$O$_3$ structure are closer to $\alpha$-Fe$_2$O$_3$ than those in the $\theta$-Al$_2$O$_3$ structure. 
Only the locations of entire peaks of Fe$_2$O$_3$ in the $\alpha$-Al$_2$O$_3$ structure were slightly shifted toward higher $2\theta$ region because of smaller lattice constants. 
However, with respect to the target XRD pattern, $S_\textrm{cos}$ of the XRD pattern of Fe$_2$O$_3$ in the $\alpha$-Al$_2$O$_3$ structure is only 24\% which is lower than that of Fe$_2$O$_3$ in the $\theta$-Al$_2$O$_3$ structure (31\%). 

\begin{figure*}[b]
 \begin{center}
  \includegraphics[width=0.58\linewidth]{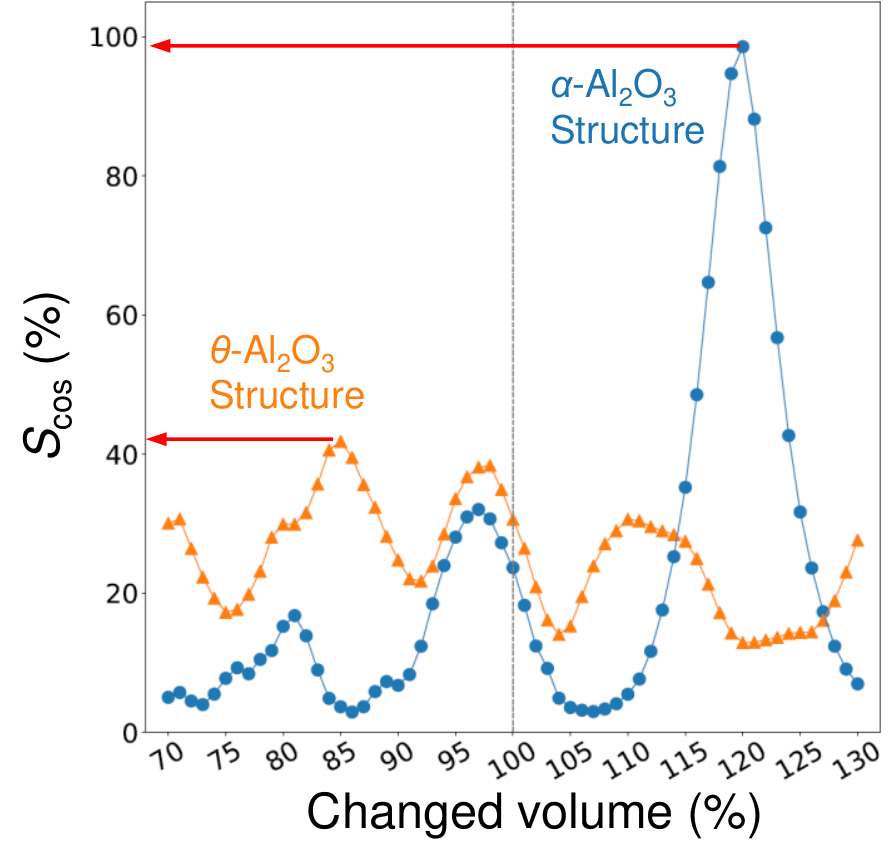}
 \caption{
 \textbf{Solution to the sensitivity problem of $S_\textrm{cos}$ by isotropic volume scanning.}
   Dependence of $S_\textrm{cos}$ of XRD patterns of Fe$_2$O$_3$ in the Al$_2$O$_3$ structures with respect to target on the isotropic volume changes. 
  Red indicators denote the highest $S_\textrm{cos}$ among the investigated volume points.
 }
\label{fig:cossimvol}
\end{center}
\end{figure*}

This inconsistency can be corrected by obtaining the highest $S_\textrm{cos}$ with the scanning along isotropic volume changes. 
Supplementary Figure \ref{fig:cossimvol} shows the dependence of the $S_\textrm{cos}$ of the XRD pattern of Fe$_2$O$_3$ in the Al$_2$O$_3$ structures with respect to the target XRD pattern on the volume change of structures. 
The $S_\textrm{cos}$ for the $\alpha$-Al$_2$O$_3$ structure drastically increased at the volume ratio greater than one and reached 99\% at a changed volume of 120\%. 
Actually, the unit cell volume of $\alpha$-Fe$_2$O$_3$ is larger than that of $\alpha$-Al$_2$O$_3$ by 20\%. 
Meanwhile, the $S_\textrm{cos}$ for the $\theta$-Al$_2$O$_3$ did not significantly change and remained below 42\% in the scanned volume change. 
Therefore, the scanning along isotropic volume change makes $S_\textrm{cos}$ to avoid a drastic drop ascribed to a small lattice parameter mismatch
and help to find a crystal structure to generate an XRD pattern similar to the target more correctly. 
Therefore, $S_\textrm{cos}$ used in this study indicates the maximum value obtained after isotropic volume scanning. 
The isotropic volume changes considered are summarized in Supplementary Table \ref{tab:parameters}.

\begin{table}[h]
\begin{center}
\begin{minipage}{\textwidth}
\caption{Parameters investigated for determining the highest $S_\textrm{cos}$.}
\label{tab:parameters}
\begin{tabular*}{\textwidth}{@{\extracolsep{\fill}}lll@{\extracolsep{\fill}}}
\toprule%
Work & Target & Considered parameter values\\
\midrule%
Determination of the & Changed  & 85, 90, 92, 94, 96, 97, 98, 99, 100, 101, 102, 103,  \\
highest $S_\textrm{cos}$ & volume (\%) & 104, 106, 108, 110, 115 \\
\midrule%
Determination of & Tolerance  & 0.001, 0.002, 0.003, 0.005, 0.007, 0.01, 0.02, 0.03,\\
space group & value & 0.05, 0.07, 0.1, 0.2, 0.3, 0.5, 0.7, 1\\
during symmetrization & & \\
\botrule
\end{tabular*}

\end{minipage}
\end{center}
\end{table}
\clearpage

\clearpage
\begin{figure*}[t]
 \begin{center}
  \includegraphics[width=0.95\linewidth]{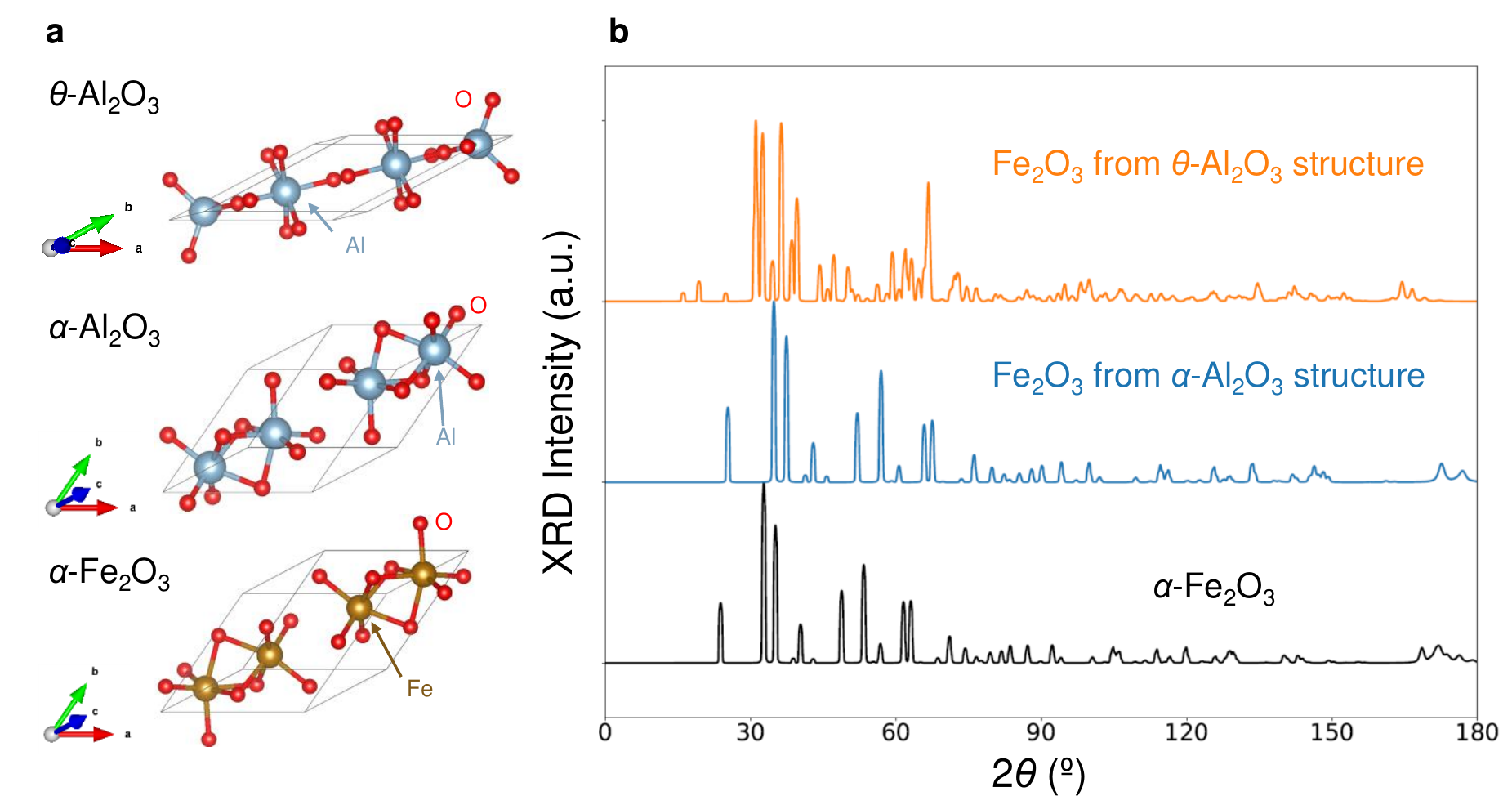}
 \caption{
 \textbf{Example of similar and different structures.}
  \textbf{a} Crystal structures of Fe$_2$O$_3$ and Al$_2$O$_3$. 
  \textbf{b} XRD patterns of $\alpha$-Fe$_2$O$_3$ structure and Fe$_2$O$_3$ in two different Al$_2$O$_3$ structures (Fe atoms occupy all Al sites). 
  The peaks were broadened with Gaussian (factor of 0.5). 
  For XRD intensity, Min-Max normalization was performed.
  }
\label{fig:al2o3xrd}
\end{center}
\end{figure*}

\clearpage

\subsection*{Comparison of performance with refinement after DB-search}\label{subsec9}

Our scheme that directly creates crystal structure by the evolutionary algorithm and crystal morphing (Evolv\&Morph) is compared with Rietveld refinement after a database(DB)-search.
In this study, the target XRD pattern was calculated from structures in Inorganic Crystal Structure Database (ICSD), which mainly records experimentally measured data.
To prepare the vanilla strategy for a comparison, close structures to the target were loaded from Materials Project Database (MPD) \cite{MPD}, which mainly contains structures obtained by first-principles calculations. 
In addition, close structures to the target from another DB including records experimentally measured data, SpringerMaterials (SM) dabatase,\cite{springer-site} were also prepared.
Then, BBO-Rietveld and symmetrization were performed after the lattice volumes with the highest $S_\textrm{cos}$ were adjusted by isotropic volume scanning.
Note that the close structures, with slightly different lattice parameters and internal coordinates of atoms, are linked in ICSD, SM, and MPD among others;
therefore, it is advantageous in finding a correct structure because it could perform refinement with a close structure.

Supplementary Table \ref{tab:vsmpd} shows the result. Except for Li$_5$BiO$_5$, the highest $S_\textrm{cos}$ from either MPD or SM obtained were $>$ 90\%; 
however, the scores were similar to or lower than those obtained by Evolv\&Morph. 
Li$_5$BiO$_5$ refined from the structures from MPD and SM exhibited only $S_\textrm{cos}$ of 66\% and 70\%, respectively. 
For this system, the optimized structure by the first-principles calculation differed from the structure in ICSD.
The target structure of Li$_5$BiO$_5$ had lattice angles of $\alpha$ = $\beta$ = 90\degree, while the refined structures from MPD or SM had lattice angles of $\alpha$ = $\beta$ = $\sim$108\degree.
The XRD pattern and crystal structure are shown in Supplementary Fig. \ref{fig:cifxrdmpd}. 
For a case where finding a close structure in DB is complex, determining a correct structure for reproducing the XRD pattern is more difficult by a naive method. 
This suggests a strong motivation that a clever inverse design method such as Evolv\&Morph is required.

The comparison test was also performed with the powder XRD pattern as the target.
The powder XRD pattern was imported from RRUFF mineral database.\cite{RRUFF,RRUFF-site}
For the vanilla strategy, the structure in ICSD was also employed. 
As written in main text, the background hinders the creation of correct structure to reproduce the target XRD pattern; therefore, the background was eliminated prior to the comparison. 
Supplementary Table \ref{tab:vs-rruff-mpd} shows the result. Similar to the case, the $S_\textrm{cos}$ score obtained by Evolv\&Morph were similar to or higher than those obtained by the vanilla strategy.

\begin{table}[ht]
\begin{center}
\begin{minipage}{1\textwidth}
\caption{Refinement result using crystal structures loaded from DB different from ICSD. 
The space groups of the structures loaded from other DB were confirmed to be the same as those of the target ones prepared from ICSD.}
\label{tab:vsmpd}

\begin{tabular*}{1.1\textwidth}{@{\extracolsep{\fill}}llcllclc@{\extracolsep{\fill}}}
\toprule%
\multicolumn{4}{@{}c@{}}{Target from ICSD} & \multicolumn{2}{@{}c@{}}{Loaded from MPD} & \multicolumn{2}{@{}c@{}}{Loaded from SM} \\
\cmidrule{1-4}\cmidrule{5-6}\cmidrule{7-8}
Materials & ICSD & Space & Crystal & MPD & Highest & SM & Highest  \\
~ & number & group & structure & number & $S_\textrm{cos}$ (\%) & number & $S_\textrm{cos}$ (\%) \\
\midrule%
Mg$_4$O$_4$ & 52026 & $Fm$–3$m$ & Cubic & 1265 & 100 & 0305005 & 100 \\
NbCu$_3$Se$_4$ & 628485 & $P$–43$m$ & Cubic & 4043 & 98.9 & NA\footnotemark[1] & NA\\
GaN & 34476 & $P$6$_3mc$ & Hexagonal & 804 & 97.5 & 0526903 & 99.8 \\
Zr$_3$Cu$_4$Si$_2$ & 26260 & $P$–62$m$ & Hexagonal & 7930 & 97.7 & 0460875 & 87.5 \\
Al$_2$O$_3$ & 9772 & $R$–3$c$ & Trigonal & 1143 & 100 & 0315064 & 99.6  \\
AlAgS$_2$ & 25356 & $P$3$m$1 & Trigonal & 7885 & 99.0 & NA & NA \\
TiAl$_3$ & 163715 & $I$4/$mmm$ & Tetragonal & 542915 & 100 & 0261474 & 98.7 \\
Zr$_2$CuSb$_3$ & 195058 & $P$–4$m$2 & Tetragonal & 16421 & 98.0 & 1008476 & 99.0 \\
Mo$_2$C & 43322 & $Pbcn$ & Orthorhombic & 1552 & 98.0 & 0260539 & 96.6	 \\
LaTaO$_4$ & 238803 & $Cmc$2$_1$ & Orthorhombic & 3998 & 90.9 & 1219799 & 98.1  \\
ZrO$_2$ & 291451 & $P$2$_1$/$c$ & Monoclinic & 2858 & 96.0 & 1638374 & 97.3  \\
Li$_5$BiO$_5$ & 203031 & $Cm$ & Monoclinic & 29365 & 65.5 & 1614623 & 70.0\\

\botrule
\end{tabular*}

\footnotetext[a]{ Not available because there are no records.}

\end{minipage}
\end{center}
\end{table}
\clearpage

\begin{sidewaystable}[ht]
\begin{center}
\begin{minipage}{1\textwidth}
\caption{Refinement result using crystal structures loaded from DB with the target of powder XRD patterns from RRUFF. 
The space groups of the structures loaded from other DB were confirmed to be the same as those of the target ones prepared from RRUFF.
Note that the background noises of the target XRD were removed.
}
\label{tab:vs-rruff-mpd}

\begin{tabular*}{1\textwidth}{@{\extracolsep{\fill}}llcllclclc@{\extracolsep{\fill}}}
\toprule%
\multicolumn{4}{@{}c@{}}{Target from RRUFF } & \multicolumn{2}{@{}c@{}}{Loaded from MPD} & \multicolumn{2}{@{}c@{}}{Loaded from SM} & \multicolumn{2}{@{}c@{}}{Loaded from ICSD} \\
\cmidrule{1-4}\cmidrule{5-6}\cmidrule{7-8}\cmidrule{9-10}
Materials & RRUFF & Space & Crystal & MPD & Highest & SM & Highest & ICSD & Highest \\
~ & number & group & structure & number & $S_\textrm{cos}$ (\%) & number & $S_\textrm{cos}$ (\%) & number & $S_\textrm{cos}$ (\%) \\
\midrule%
ZnAl$_2$O$_4$ & R040027 & $Fd$–3$m$ & Cubic & 2908 & 97.8 & 0381992 & 98.9 & 196109 & 98.7 \\
Al$_2$O$_3$ & R040096 & $R$–3$c$ & Trigonal & 1143 & 98.5 & 0315064 & 96.6 & 9772 & 96.4  \\
CaTiO$_3$ & R050456 & $Pnma$ & Orthorhombic & 4019 & 97.8 & 0381921 & 96.6 & 185443 & 83.4  \\
ZrO$_2$ & R060016 & $P$2$_1$/$c$ & Monoclinic & 2858 & 89.2 & 1638374 & 95.6 & 291451 & 94.4 \\

\botrule
\end{tabular*}

\end{minipage}
\end{center}
\end{sidewaystable}
\clearpage


\begin{figure*}[htp]
 \begin{center}
  \includegraphics[width=0.90\linewidth]{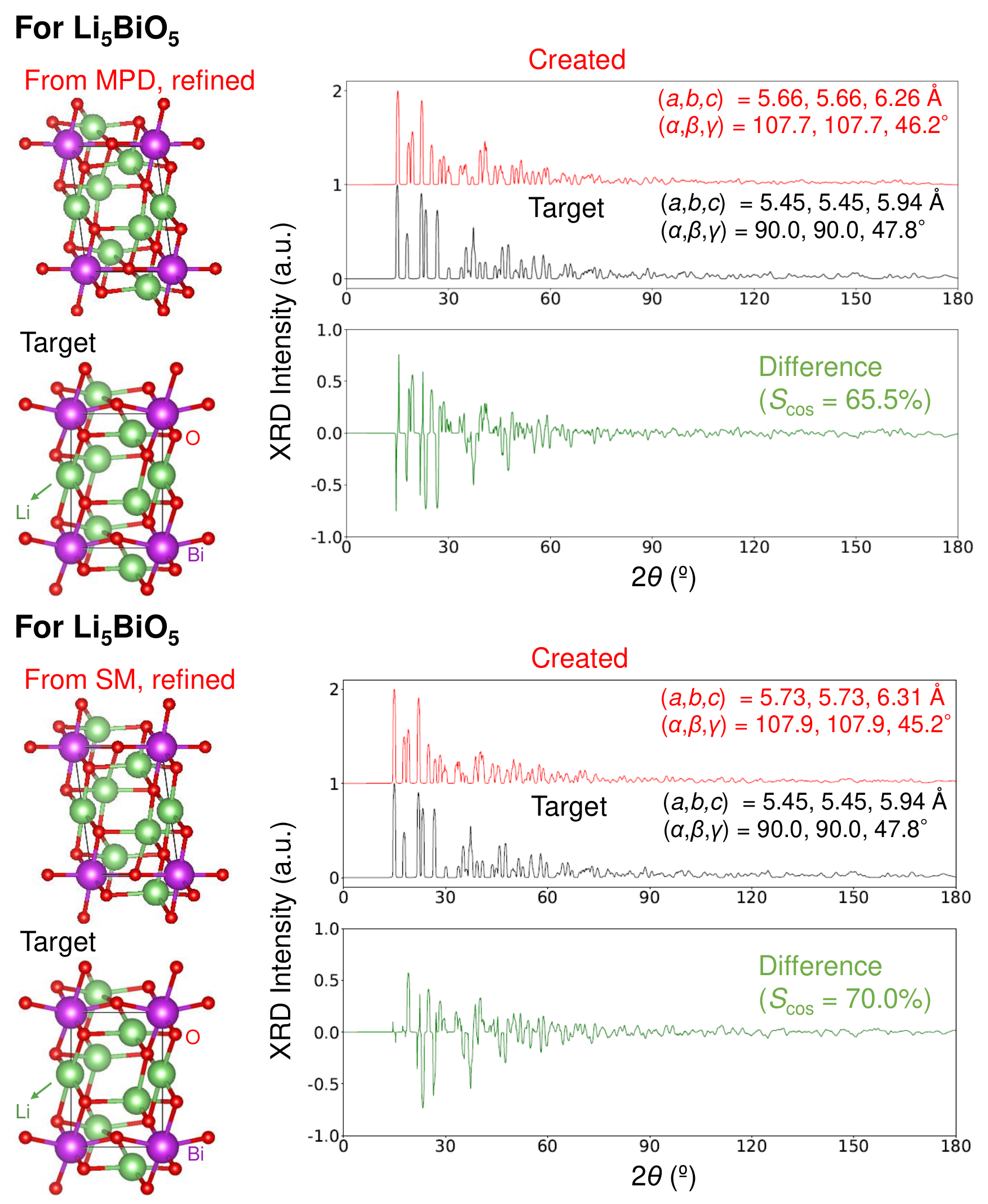}
 \caption{
 \textbf{Comparison of the target and refined structure loaded from other DB.}
   (Left panel) Comparison of the refined structure loaded from other DB and the target structure.
   Li$_5$BiO$_5$ system with $S_\textrm{cos}$ $<$ 70\% is shown as the case where the refinements using crystal structures loaded from other DB such as MPD and SM are not successful.
   (Right panel) Comparison of XRD pattern from the refined crystal structure loaded from other DB and target XRD pattern 
   and their difference.
   XRD patterns were rescaled by Min–Max normalization. 
   XRD peaks were broadened with Gaussian smearing of $0.5^{\circ}$. ($a$, $b$, $c$) and ($\alpha$, $\beta$, $\gamma$) written in the boxes are the lengths and angles of unitcell, respectively. 
}
\label{fig:cifxrdmpd}
\end{center}
\end{figure*}
\clearpage

\subsection*{Target and created crystal structures, and their XRD patterns}\label{ssubsec2}
\vspace{1em}

\begin{figure*}[!h]
 \begin{center}
  \includegraphics[width=0.87\linewidth]{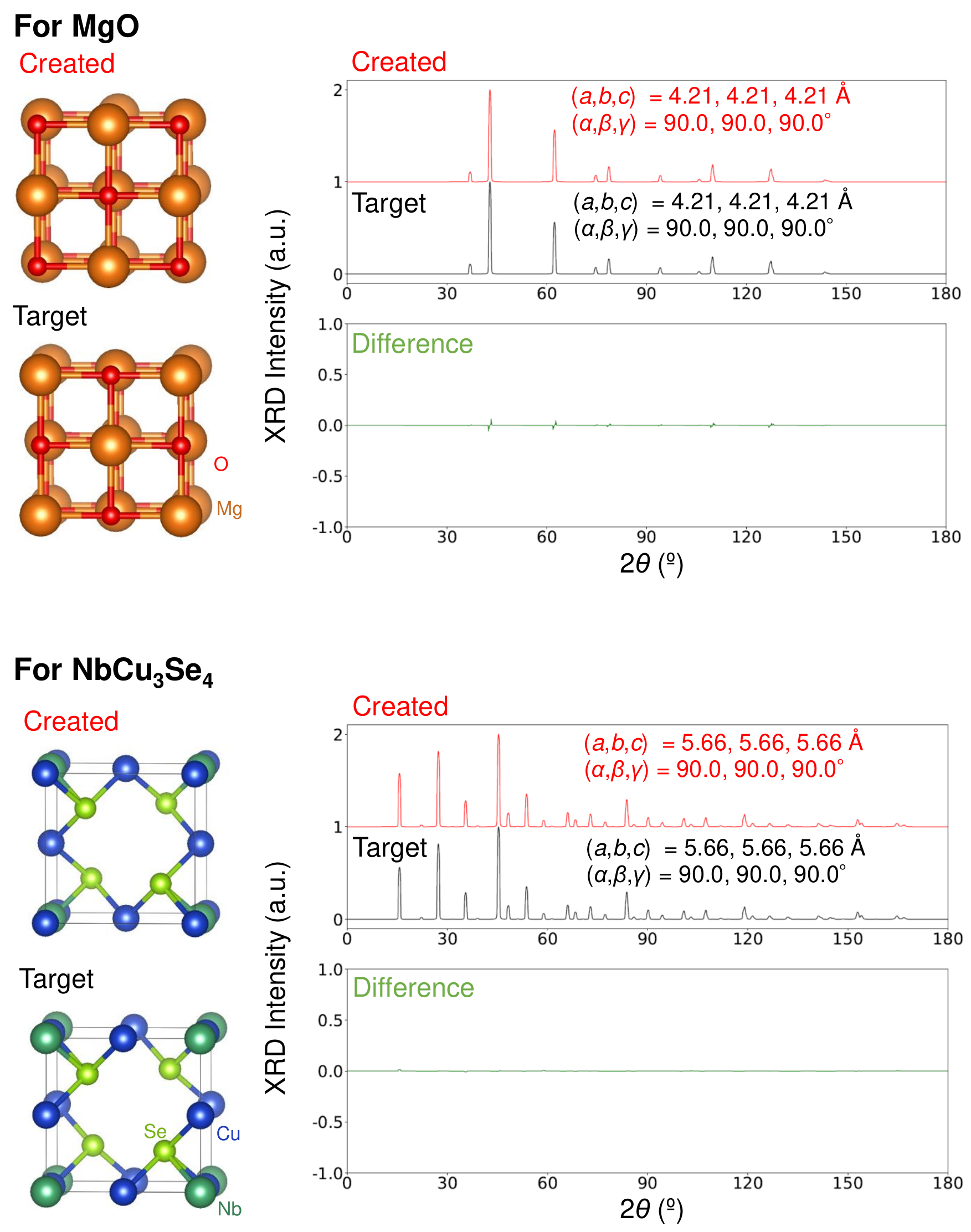}
 \caption{
 \textbf{Created structures and their XRD patterns obtained in this study when target XRD pattern is generated by simulation using structures in ICSD.}
  (Left panel) Comparison of the created and the target crystal structures. 
  (Right panel) Comparison of XRD patterns from the created and target structures, and their difference. XRD patterns were rescaled by Min–Max normalization. 
  XRD peaks were broadened with Gaussian smearing of $0.5^{\circ}$. ($a$, $b$, $c$) and ($\alpha$, $\beta$, $\gamma$) written in the boxes are the lengths and angles of unitcell, respectively.
  }
\label{fig:cifxrd1}
\end{center}
\end{figure*}

\begin{figure*}[htp]
\addtocounter{figure}{-1}
 \begin{center}
  \includegraphics[width=0.87\linewidth]{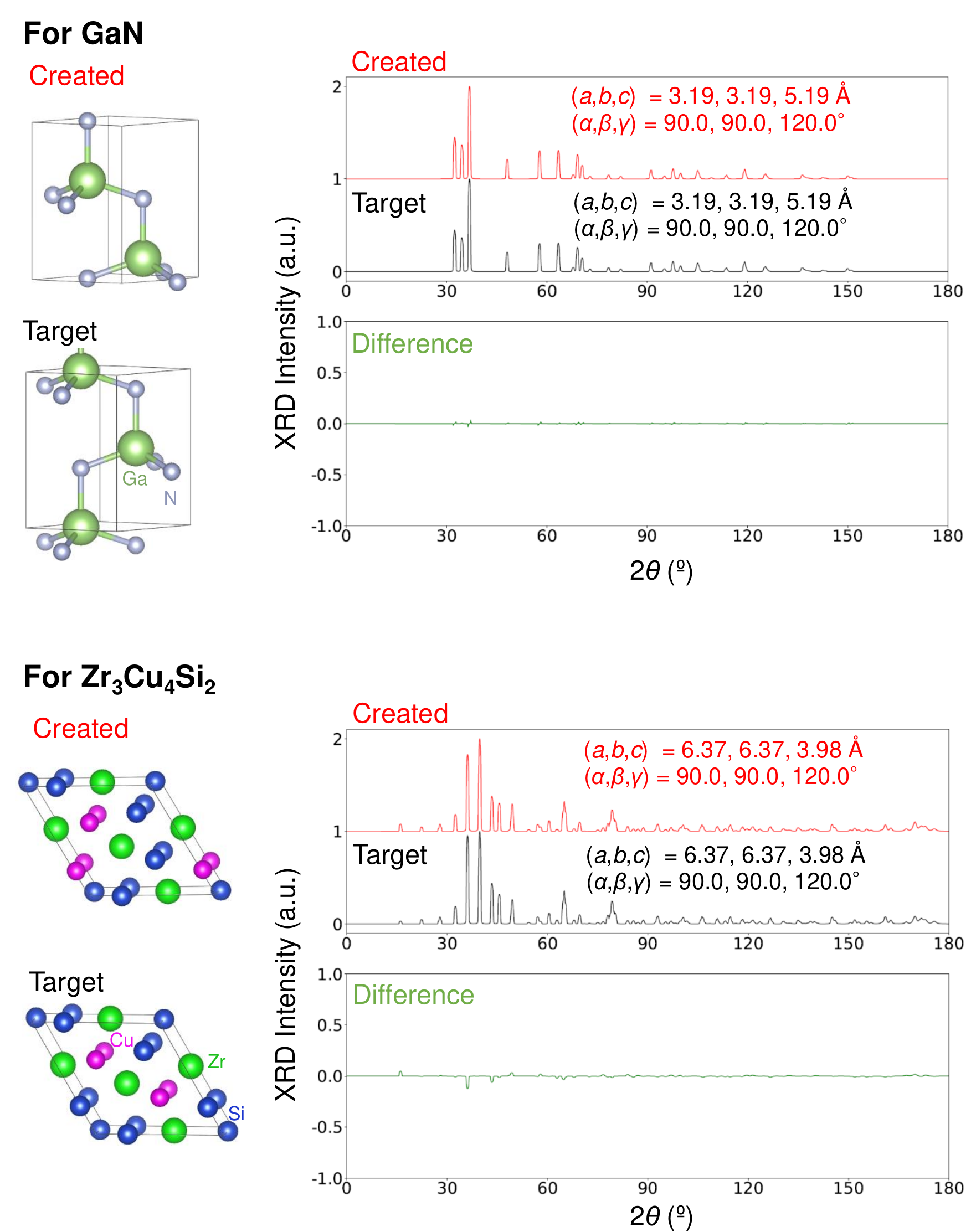}
 \caption{
  (Continued.)
   }
\label{fig:cifxrd2}
\end{center}
\end{figure*}
\clearpage
\begin{figure*}[htp]
\addtocounter{figure}{-1}
 \begin{center}
  \includegraphics[width=0.87\linewidth]{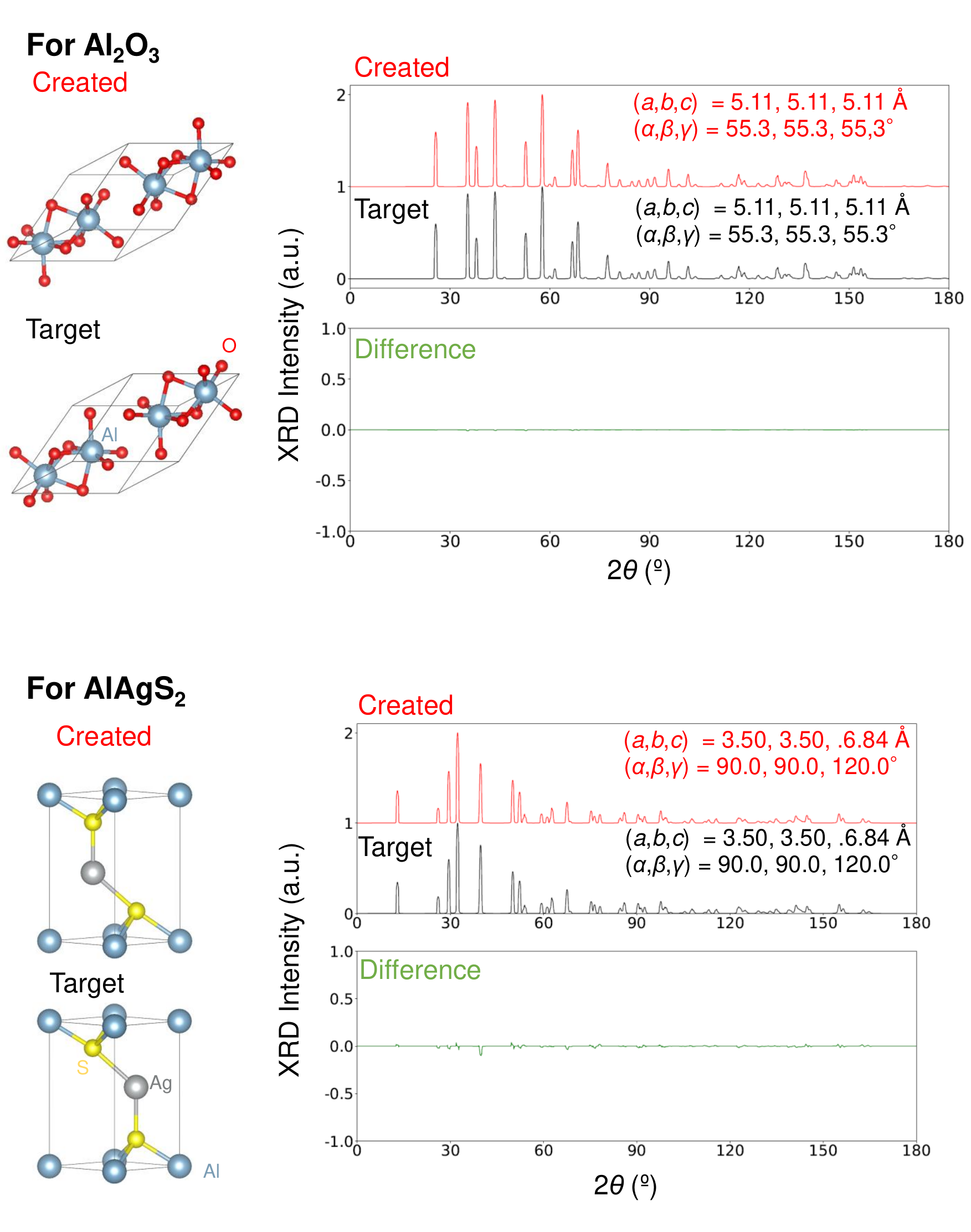}
 \caption{
  (Continued.)
 }
\label{fig:cifxrd3}
\end{center}
\end{figure*}
\clearpage
\begin{figure*}[htp]
\addtocounter{figure}{-1}
 \begin{center}
  \includegraphics[width=0.87\linewidth]{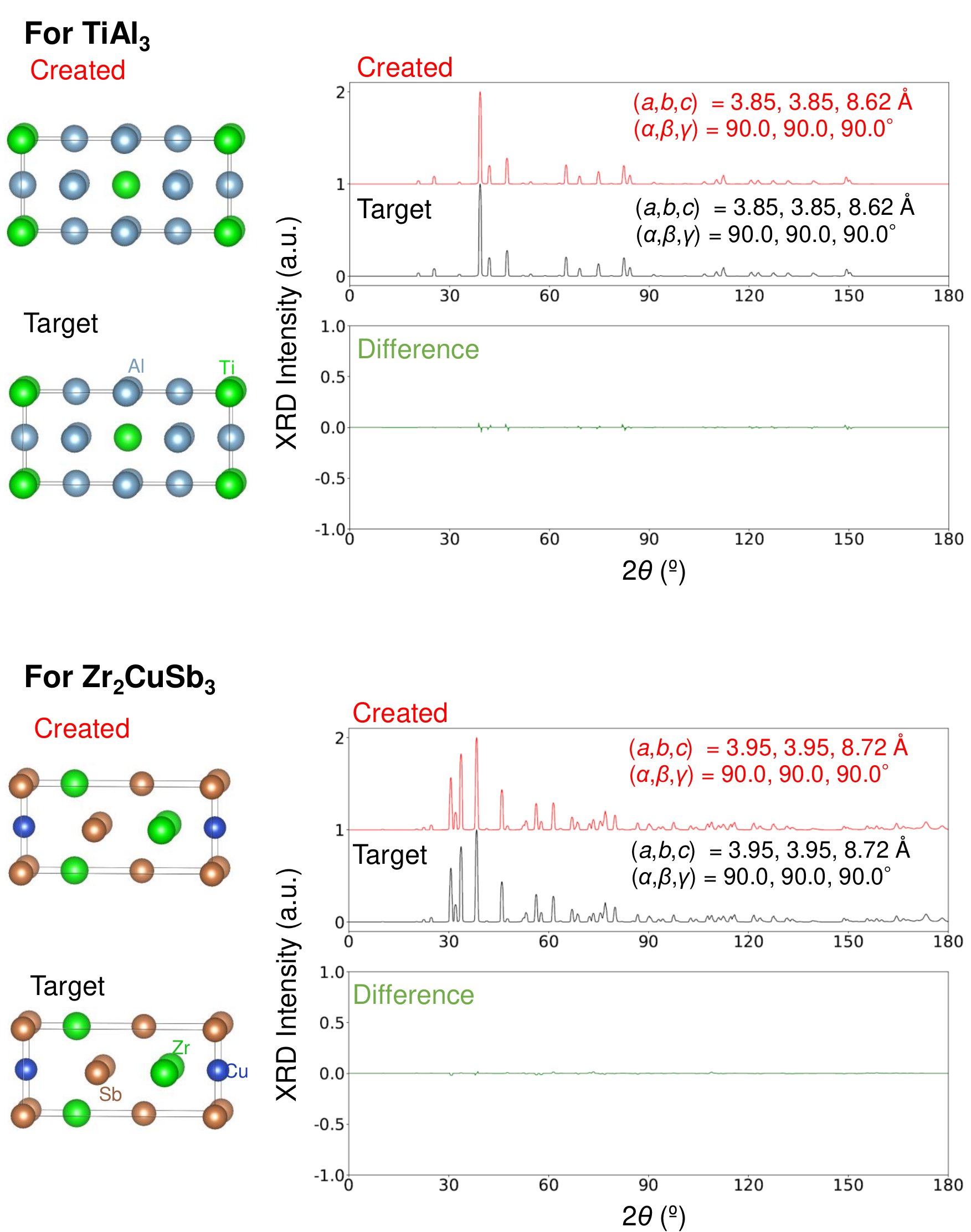}
 \caption{
  (Continued.)
  }
\label{fig:cifxrd4}
\end{center}
\end{figure*}
\clearpage
\begin{figure*}[htp]
\addtocounter{figure}{-1}
 \begin{center}
  \includegraphics[width=0.87\linewidth]{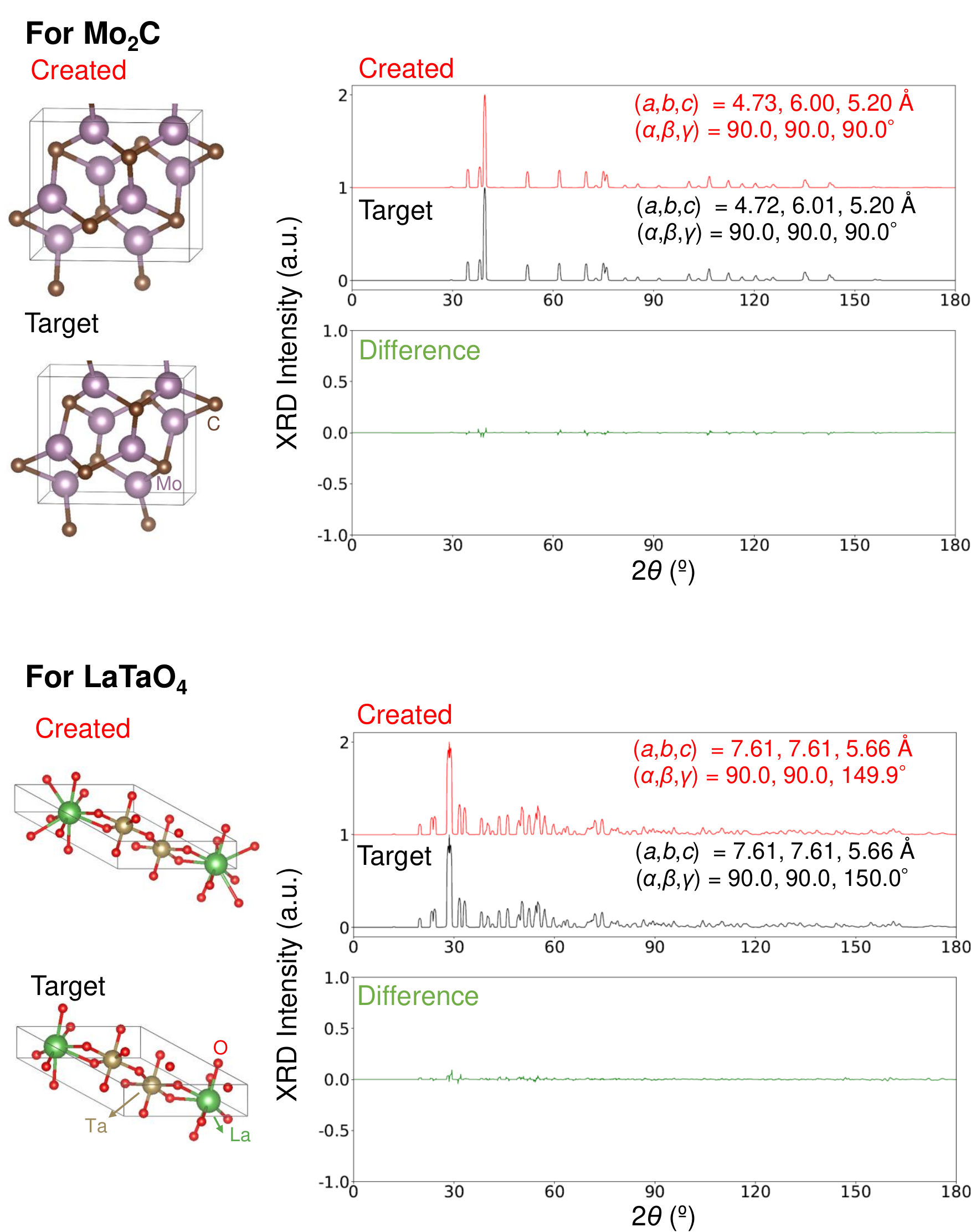}
 \caption{
  (Continued.)
 }
\label{fig:cifxrd5}
\end{center}
\end{figure*}
\clearpage
\begin{figure*}[htp]
\addtocounter{figure}{-1}
 \begin{center}
  \includegraphics[width=0.87\linewidth]{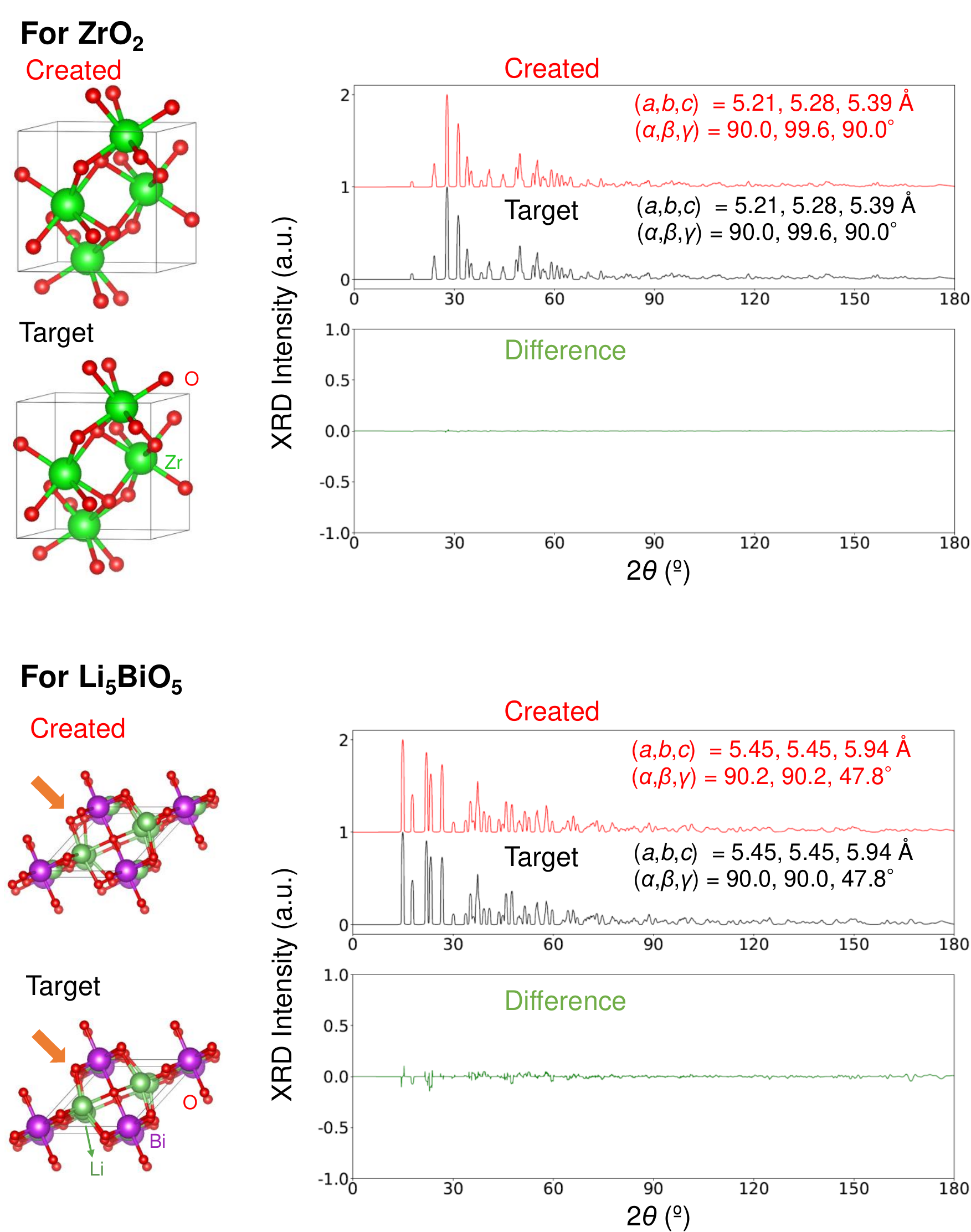}
 \caption{
  (Continued.) 
  For Li$_5$BiO$_5$ system, an orange indicator shows a slightly different part of the created and target crystal structures.
 }
\label{fig:cifxrd6}
\end{center}
\end{figure*}
\clearpage
\begin{figure*}[!h]
 \begin{center}
  \includegraphics[width=0.87\linewidth]{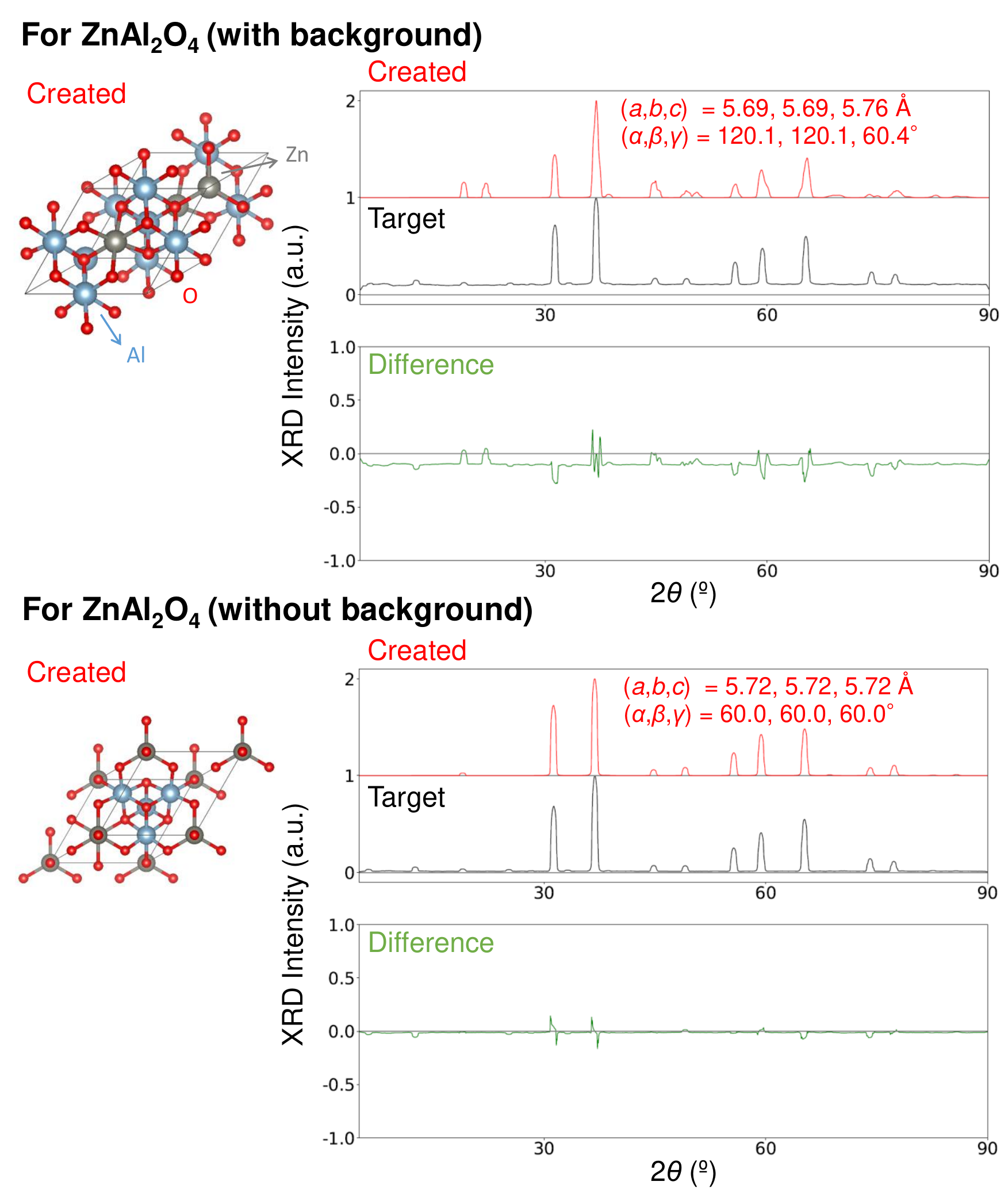}
 \caption{
 \textbf{Created structures and their XRD patterns obtained in this study when the powder XRD pattern is used as the target.}
 Comparison of XRD patterns of created structures and target XRD patterns.
 XRD patterns were rescaled by the Min–Max normalization.
 XRD peaks were broadened with Gaussian smearing of $0.5^{\circ}$. ($a$, $b$, $c$) and ($\alpha$, $\beta$, $\gamma$) written in the boxes are the lengths and angles of unitcell, respectively.
 (Upper panel) The background of the target XRD pattern is kept. (Lower panel) The background of the target XRD pattern is removed.
}
\label{fig:cifxrdexp1}
\end{center}
\end{figure*}
\clearpage
\begin{figure*}[htp]
\addtocounter{figure}{-1}
 \begin{center}
  \includegraphics[width=0.87\linewidth]{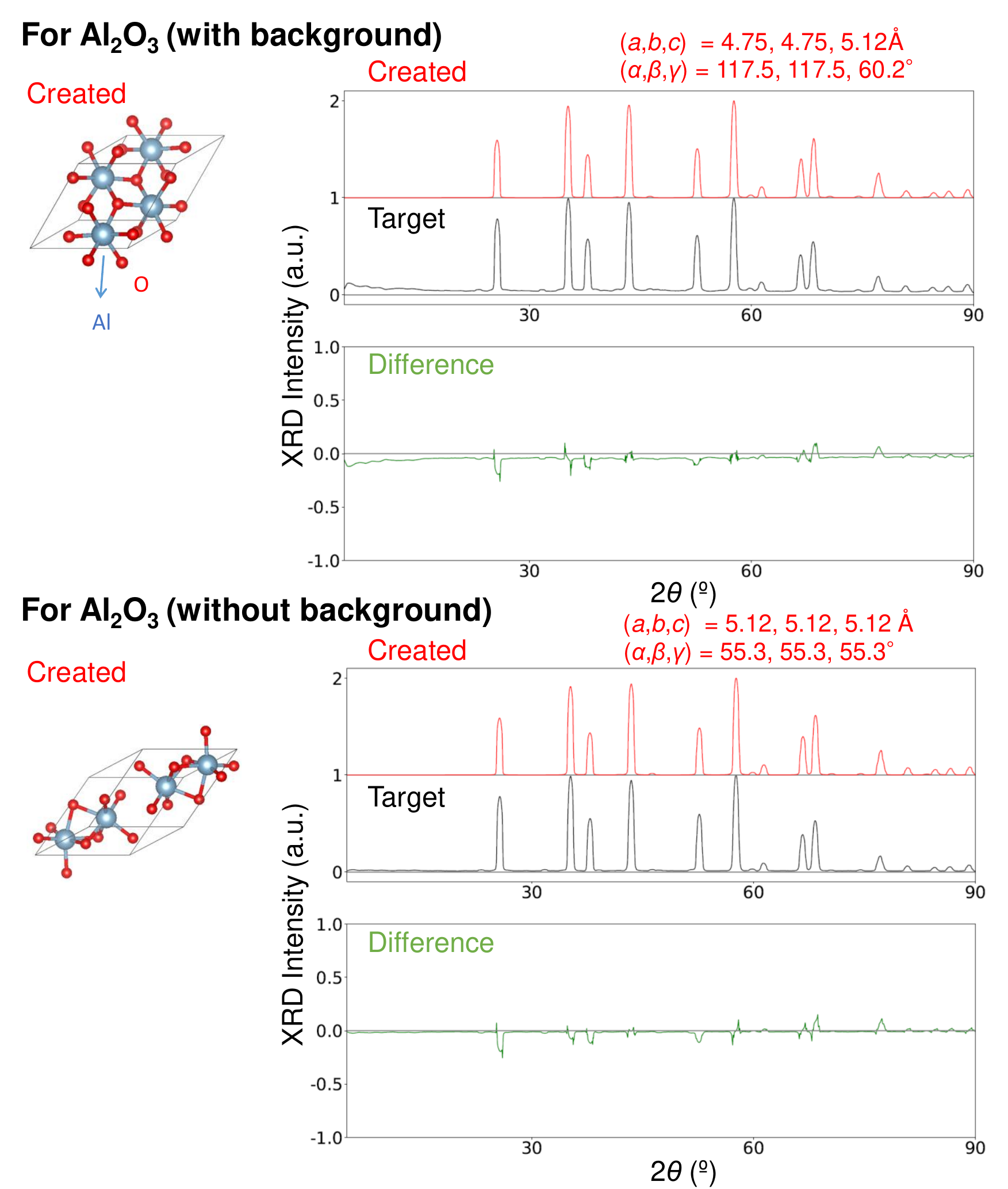}
 \caption{
  (Continued.)
 }
\label{fig:cifxrdexp2}
\end{center}
\end{figure*}
\clearpage
\begin{figure*}[htp]
\addtocounter{figure}{-1}
 \begin{center}
  \includegraphics[width=0.87\linewidth]{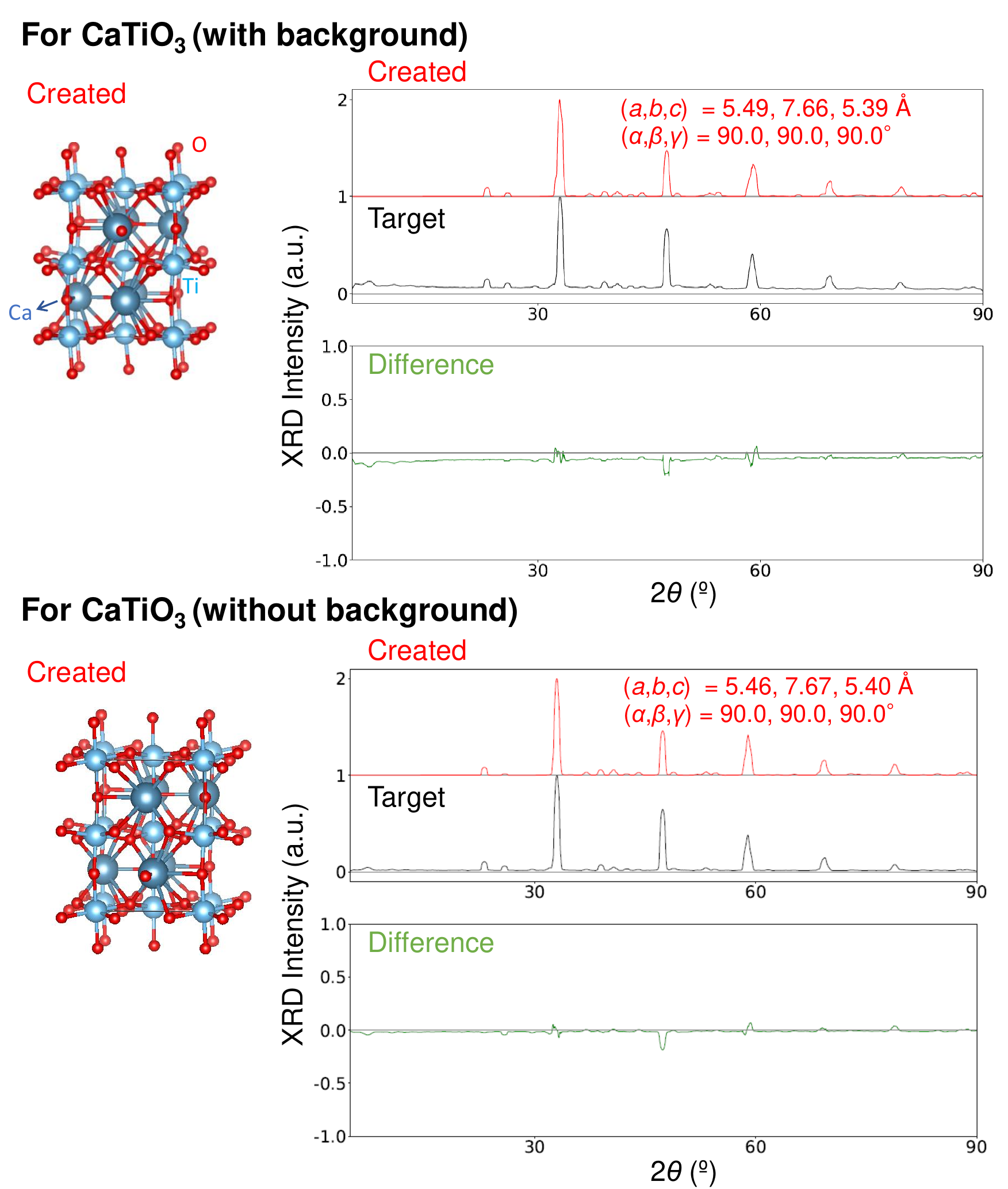}
 \caption{
  (Continued.)
 }
\label{fig:cifxrdexp3}
\end{center}
\end{figure*}
\clearpage
\begin{figure*}[htp]
\addtocounter{figure}{-1}
 \begin{center}
  \includegraphics[width=0.87\linewidth]{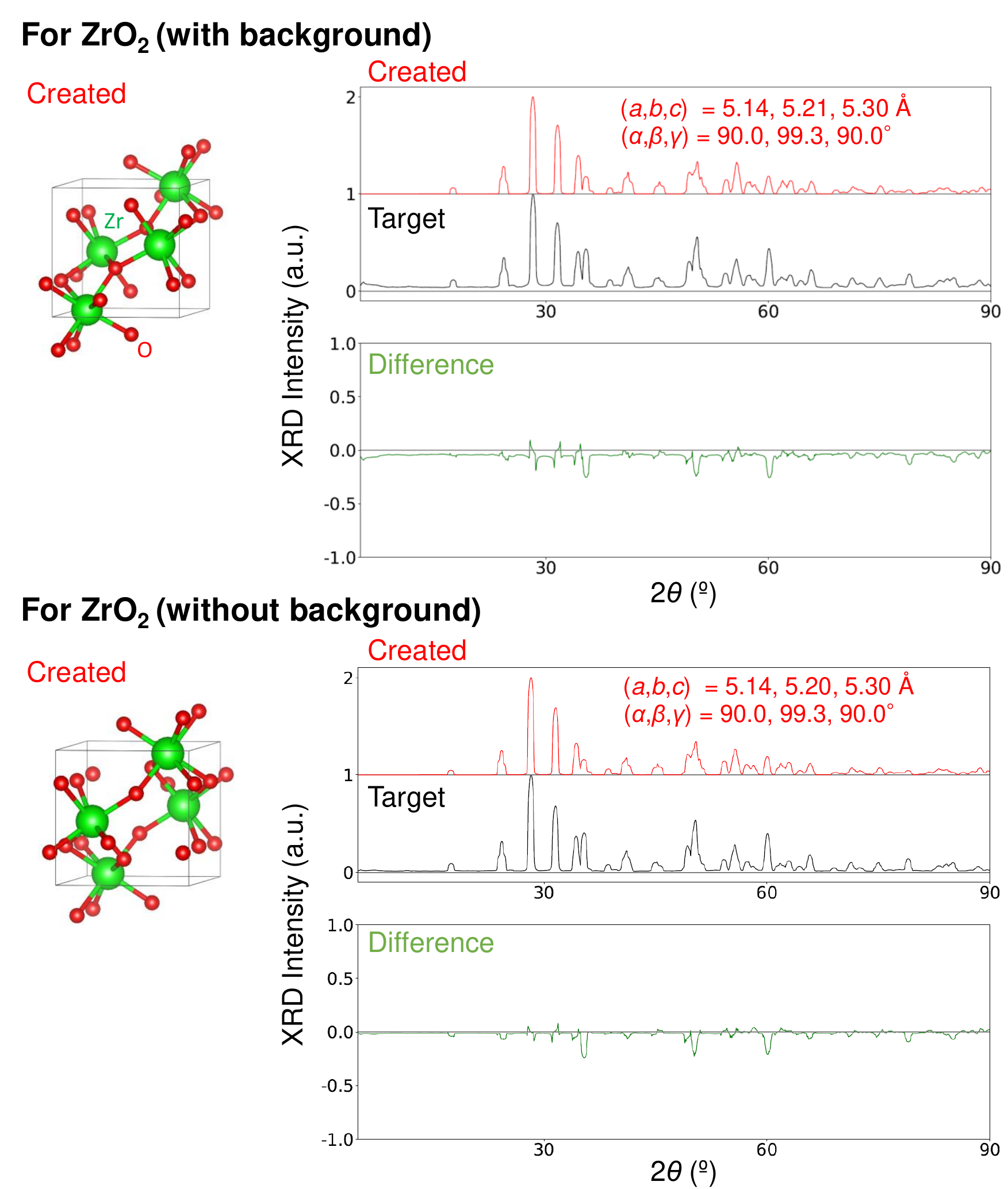}
 \caption{
  (Continued.)
 }
\label{fig:cifxrdexp4}
\end{center}
\end{figure*}
\clearpage

\end{document}